\documentclass{article}
\usepackage[a4paper, margin=0.8in]{geometry}
\usepackage{amsmath}
\usepackage{url}

\usepackage{breakurl}
\usepackage{hyperref}
\usepackage{hyperxmp} 
\usepackage{fix-cm}
\usepackage{booktabs,caption}
\usepackage[flushleft]{threeparttable}
\usepackage{supertabular}
\usepackage{bbding}
\usepackage{wasysym}
\usepackage{pgf,tikz,pgfplots}
\usepackage{rotating,booktabs,chemformula}
\usepackage{adjustbox}
\usepackage{float}
\usepackage{booktabs, makecell, longtable}
\usepackage{lipsum} 
\usepackage{rotating} 
\usepackage{multirow}
\usepackage{enumitem}
\usepackage{amssymb}
\usepackage{pifont}
\usepackage[normalem]{ulem}
\usepackage{tikz}
\usepackage{longtable}
\usepackage{anyfontsize}
\usepackage{authblk}
\usepackage{array}

\makeatletter
\def\bstctlcite{\@ifnextchar[{\@bstctlcite}{\@bstctlcite[@auxout]}}
\def\@bstctlcite[#1]#2{\@bsphack
  \@for\@citeb:=#2\do{%
    \edef\@citeb{\expandafter\@firstofone\@citeb}%
    \if@filesw\immediate\write\csname #1\endcsname{\string\citation{\@citeb}}\fi}%
  \@esphack}
\makeatother 

\usetikzlibrary{shapes.geometric, arrows}
\pgfplotsset{compat=1.18}
\useunder{\uline}{\ul}{}
\newcommand{\cmark}{\ding{51}}%
\newcommand{\xmark}{\ding{55}}%
\newcommand*\emptycirc[1][0.7ex]{\tikz\draw (0,0) circle (#1);} 
\newcommand*\halfcirc[1][0.7ex]{%
  \begin{tikzpicture}
  \draw[fill] (0,0)-- (90:#1) arc (90:270:#1) -- cycle ;
  \draw (0,0) circle (#1);
  \end{tikzpicture}}
\newcommand*\fullcirc[1][0.8ex]{\tikz\fill (0,0) circle (#1);} 
\usepackage[numbers]{natbib}

\def\tsc#1{\csdef{#1}{\textsc{\lowercase{#1}}\xspace}}
\tsc{WGM}
\tsc{QE}

\title{Unfolding Challenges in Securing and Regulating Unmanned Air Vehicles}

\author{Sonali Rout\thanks{\texttt{sonalirout105@gmail.com}}}
\author{Vireshwar Kumar\thanks{\texttt{viresh@cse.iitd.ac.in}}}
\affil{Indian Institute of Technology Delhi, New Delhi, India}

\date{} 
\begin{document}
\maketitle
\begin{abstract}
Unmanned Aerial Vehicles (UAVs) or drones are being introduced in a wide range of commercial applications. This has also made them prime targets of attackers who compromise their fundamental security properties, including confidentiality, integrity, and availability. As researchers discover novel threat vectors in UAVs, the government and industry are increasingly concerned about their limited ability to secure and regulate UAVs and their usage. With the aim of unfolding a path for a large-scale commercial UAV network deployment, we conduct a comprehensive state-of-the-art study and examine the prevailing security challenges. Unlike the prior art, we focus on uncovering the research gaps that must be addressed to enforce security policy regulations in civilian off-the-shelf drone systems. To that end, we first examine the known security threats to UAVs based on their impact and effectiveness. We then analyze existing countermeasures to prevent, detect, and respond to these threats in terms of security and performance overhead. We further outline the future research directions for securing UAVs. Finally, we establish the fundamental requirements and highlight critical research challenges in introducing a regulatory entity to achieve a secure and regulated UAV network.
\end{abstract}

\section{Introduction} 
\label{sec1_Introduction}
We refer to a drone as a flying machine, Unmanned Aerial Vehicle (UAV), Unmanned Aircraft System (UAS), or Remotely Piloted Aerial System (RPAS), as this aircraft operates remotely with human supervision rather than the direct presence of a human on itself~\cite{Drones}. In recent years, drones have been attracting the focus of researchers as their features, such as low cost, high maneuverability, and easy access to remote areas, enable them to be the perfect candidates for a plethora of applications in various fields. These include disaster assessment and management~\cite{daud2022applications}, agriculture \cite{shah2023application}, search and rescue~\cite{bassolillo2023distributed}, healthcare supply transportation to remote locations~\cite{amirsahami2023hierarchical}, etc. Specifically, for short-range delivery in densely populated areas, drones can be widely used~\cite{kim2020drone}. For instance, they have been utilized in Germany to deliver goods purchased in online stores since 2014~\cite{bryan2014drone}. In Rwanda, they have been used to quickly deliver blood to hospitals~\cite{nisingizwe2022effect}. Amazon has been utilizing them for its ultra-fast delivery in Italy, the UK, and the USA~\cite{Amazondrones}.

In this paper, we limit our study to civilian off-the-shelf drones~\cite{hassanalian2017classifications}. These are ready-to-use drones that can be bought easily and can be used with low technical expertise~\cite{humpe2020bridge}. Unfortunately, due to their easy availability and abundance in the market, off-the-shelf drones can also be used in various malevolent ways ~\cite{Drug_delivery_by_drones}, e.g., conducting surveillance in a restricted area and dropping an explosive payload in a sensitive area. They can also be compromised by malicious attackers and then exploited in nefarious ways, e.g., causing a collision with a benign aircraft. Due to these potential threats, the stakeholders, including the governments and the industry, are concerned about deploying drones for large-scale commercial applications, e.g., in delivery services. 

To mitigate the possibility of potential harm by such drones, some countries have proposed regulations that must be followed by drones~\cite{bassi2019european}. For instance, there could be restrictions in flying drones near the airport and critical infrastructures. Such restrictions are typically pre-loaded on the drones by their manufacturers. In the real world, the enforcement agencies also announce dynamic restrictions on the usage of drones, e.g., during events that attract large crowds and during the movement of sensitive personnel. Unfortunately, there are limited methods to enforce these dynamic regulations if a drone fails to follow them, either due to the malicious intent of its user or due to compromise by an attacker.

In this paper, we present a systematic analysis of the state-of-the-art threats and countermeasures considering different attack surfaces, such as hardware, software, communication module, and sensors. We note that while the existing studies have investigated security aspects of drones, very few works have focused on the technical advancements required for security policy regulation and enforcement~\cite{bhat2024autonomous}. Unlike the prior art, we explore and document the possible threats as well as their available countermeasures in drones from the regulatory and enforcement perspective, and focus on their impacts on security and performance on drones.

Based on the obtained insights, we explore the answer to a fundamental question: how could a \textit{regulator/enforcer} be introduced in a UAV network to address the issue of enforcing security policies? We highlight that existing enforcement mechanisms lack hardware and software support to ensure secure and controlled drone operations, e.g., preventing drones from entering dynamically-defined restricted airspace. Our study examines these limitations and points out the need for policy-driven enforcement and regulatory strategies in drone networks. By highlighting these gaps, we aim to encourage the research community to work towards regulatory frameworks that enhance the overall usability of drones. Our key contributions are as follows.  

\begin{itemize}[noitemsep]
    \item We present a systematic taxonomy to analyze the threats across different components of drone systems and their potential countermeasures in terms of their impact on security and performance. We also present a novel classification of attacks and corresponding countermeasures that highlights their feasibility and usability.  
    
    \item We outline the technical challenges and future research directions for enhancing UAV security through robust regulation frameworks and their effective enforcement.  
\end{itemize}

\begin{figure*}[ht]
    \centering
    \includegraphics[width=1 \linewidth]{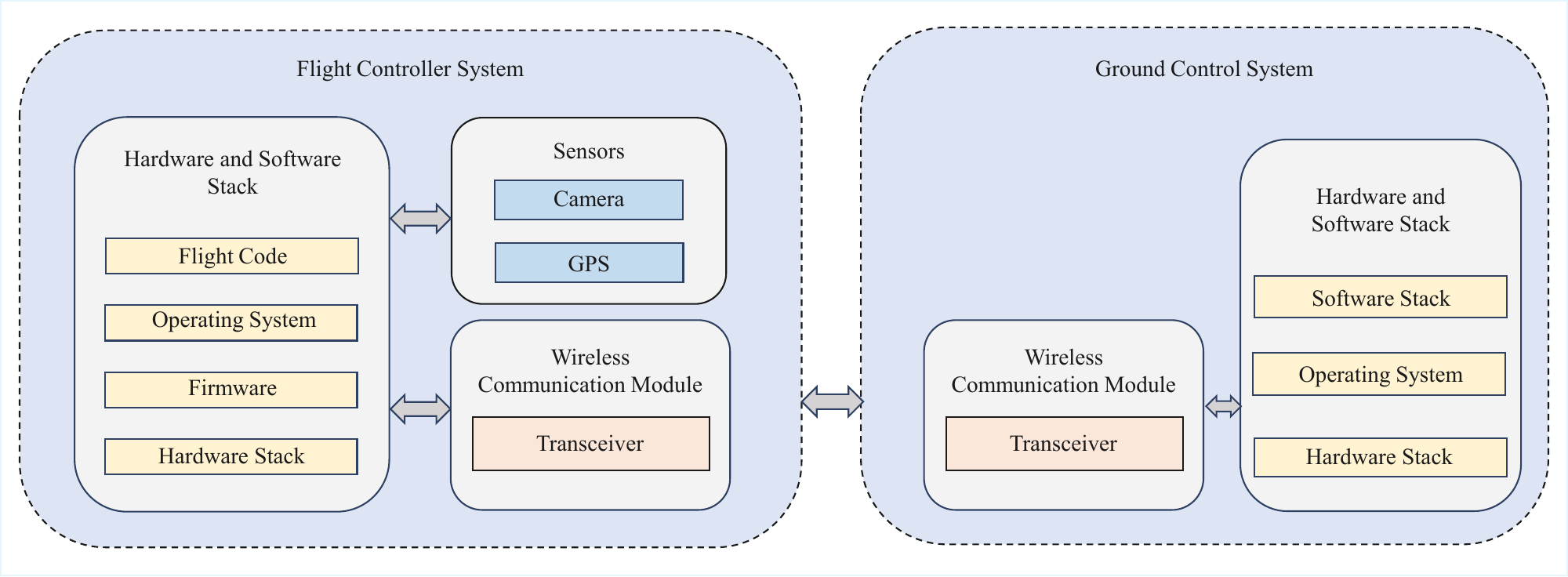}
    \caption{Illustrative architecture of a typical UAV system consisting of a flight controller and ground control with their software, hardware, and communication modules.}
    \label{drone_arch}
\end{figure*}

The remainder of this paper is organized as follows. Section~\ref{sec2_Drone_System_Architecture} outlines the architecture of a drone system, detailing its various components. Section~\ref{sec3_Limitations_of_Existing_Literature_Surveys} reviews the state-of-the-art surveys on the security challenges associated with these systems and differentiates our work from them. Section~\ref{sec4_Systematization_of_Attacks} and \ref{sec5_Systematization_of_Countermeasures} discuss a detailed taxonomy of existing attacks and countermeasures, respectively. Section~\ref{sec7_Research_Directions} outlines the research directions that could be pursued to fill the gap in the existing literature and make the drones robust against known attacks. Furthermore, Section~\ref{sec6_Security_Policies_for_Regulating_Drones} explores the feasibility of policy enforcement for regulating UAV networks. Finally, Section~\ref{sec8_Conclusion} concludes the paper with a summary of the key insights and their implications in the security and policy-guided regulation of UAVs.
\section{Drone System Architecture} 
\label{sec2_Drone_System_Architecture}

Here, we provide a brief description of the drone's system architecture. This will subsequently facilitate the detailed discussion of the threats and countermeasures in a drone's system architecture in the next sections. Figure~\ref{drone_arch} represents the system architecture of a typical drone. The components of a specific drone can vary depending on its defined purpose and manufacturer; however, the basic functional components remain similar, as discussed below. 

As shown in Figure~\ref{drone_arch}, a typical UAV system consists of a {\em flight controller system (FCS)}, which includes the hardware stack (e.g., CPU and memory), the software stack (e.g., application code, firmware, middleware, and operating system), sensors (e.g., camera and LiDAR), actuators, and a wireless {\em communication module}. The UAV system is controlled and managed by the {\em Ground Controller System (GCS)} that comprises a hardware and software stack, facilitating communication with the UAV via transceivers through its communication module. Below, we discuss these modules in detail.

\subsection{Flight Controller} 

The flight controller is the main component of a UAV system responsible for communicating with the ground control, sensing the environment, and actuating the different components. Its functionality can be discussed by diving into its software and hardware parts. The software stack of the flight controller consists of the operating system, the middleware, and the firmware. The hardware stack consists of the processor and storage. Other hardware parts include the sensors (e.g., telemetry unit and GPS receiver) and actuators. Popular boards, such as Raspberry Pi~\cite{benhadhria2021vagadrone}, Pixhawk Series~\cite{meier2011pixhawk}, and Intel Aero Compute board, can be utilized for realizing flight controllers in custom-made drones.

\subsubsection{Software Stack}

\textbf{Operating System:} The first layer of the software stack of the flight controller board is the operating system. The operating system provides the abstractions and services required to support the execution of flight control software, sensor integration, communication protocols, and other critical functions. It acts as the basis for the software stack running on a drone. Developing an operating system tailored for UAVs necessitates tackling issues including real-time reaction, limited resources, and the requirement for resilience and dependability in constantly changing flying conditions.
\\
\noindent
\textbf{Firmware:} A drone's firmware acts as a link between its hardware and the higher-level software programs that manage its functions. It is responsible for carrying out flight control algorithms, controlling hardware resources, and enabling execution of commands received from the ground control systems. Therefore, any flaws or vulnerabilities in the firmware could have a negative impact on the drone's performance, safety, and security.
\\
\noindent
\textbf{Middleware:} In a typical UAV system, different middleware is required for different purposes. In a drone's software stack, middleware is the interface that joins the various software components. It facilitates smooth coordination, data exchange, and communication between components like mission planning software, flight control algorithms, sensor integration modules, and communication protocols. Addressing issues like fault tolerance, scalability, interoperability, and real-time response in dynamic flying situations is part of designing middleware for UAVs. The middleware also plays a vital role in maintaining drone security.
\subsubsection{Hardware Stack}
\noindent
\textbf{Micro-Controller Unit (MCU):} The MCU serves as the brain of the UAV, executing control algorithms and processing data from sensors. It is responsible for managing the flight dynamics and ensuring stability during operation. The MCU's performance directly impacts the UAV's responsiveness and precision~\cite{ebeid2018survey}.

\noindent
\textbf{Storage:} Storage in UAVs typically involves onboard memory or external devices like micro SD cards. This storage is used for logging flight data, storing waypoints, and saving telemetry information. Efficient data storage is vital for post-flight analysis and troubleshooting.

\subsubsection{Other Hardware Components}
\noindent
\textbf{Sensors:} Sensors provide essential data for navigation, stability, and control of the UAV. Common sensors include gyroscopes, accelerometers, magnetometers, GPS modules, and barometers. These sensors continuously monitor the UAV's orientation, position, and altitude, feeding data to the MCU for real-time adjustments.

\noindent
\textbf{Actuators:} Actuators in UAVs include motors and servos, which convert electrical signals into physical movement. These components control the drone's propellers and control surfaces, enabling it to maneuver in the air. The proper functioning of actuators is crucial for executing flight commands accurately.

\noindent
\textbf{Connections and Ports:} The flight controller unit features multiple connections and ports, including motor output ports, receiver input ports, and USB/Serial UART ports. These interfaces enable communication with various components and peripherals, supporting the integration of additional sensors and actuators.

\subsection{Ground Control System}

The ground control system (GCS) module is used to control the drone from the ground. GCS uses a communication link to connect to the flight controller. GCS has a user interface, its own software stack, and an operating system. The operator interacts with the drone through the GCS, which provides mission planning, monitoring, and control functions. The software programs running on the GCS allow the operator to set up mission parameters like speed and altitude, organize complex flight missions, and designate waypoints. In addition, real-time telemetry data, which includes the drone's position, battery life, sensor readings, and other vital data for safe and effective flight operations is collected by the GCS software.

Furthermore, GCS makes it easier to manage the drone's flight characteristics, such as landing, taking off, navigating to waypoints, and using manual flight control. It also allows the operator to control onboard payloads, including cameras and sensors, facilitating operations like payload deployment, video recording, and picture taking. Well-known GCS programs include Mission Planner, QGroundControl, DJI GO, and Pix4Dcapture.

\subsection{Communication Link}
The drone and the ground control station are equipped with transceivers to facilitate a wireless connection link for dependable and effective communication between them. At the GCS, an operator can perform mission-critical jobs through this communication link, e.g., stream live video feeds, get telemetry data, and remotely operate the drone. The communication link typically facilitates the execution of a security protocol (e.g., Mavlink~\cite{koubaa2019micro}) to provide confidentiality and authentication for messages communicated wirelessly between the GCS and the drone.

Drone systems use various communication link types, each with unique benefits and drawbacks based on variables including range, data bandwidth, and ambient circumstances~\cite{sharma2020communication}. Radio frequency (RF) communication is a popular communication link that uses radio waves to connect the drone and base station. Different frequency bands, such as 2.4 GHz, 5.8 GHz, and UHF (Ultra High Frequency), are available for RF communication, providing varying transmission rates and ranges. 

Drones can also use cellular communication to take advantage of the current cellular infrastructure for long-range communication. This technology allows for high-bandwidth data transmission and expanded coverage beyond line-of-sight. WiFi offers minimal latency and fast data rates within a restricted range, making it a popular choice for short-range communication between drones and adjacent ground stations or mobile devices. Conversely, mesh networks allow drones to establish ad hoc networks and interact with one another, improving communication resilience and facilitating cooperative operations. Satellite communication, which uses satellite networks to facilitate long-range and beyond-line-of-sight (BLOS) communication, is another popular kind of communication. Satellite communication offers global coverage, which is especially helpful in isolated or difficult-to-reach locations where regular radio frequency communication could be restricted.
\begin{table}[htbp]
\scriptsize
\caption{Comprehensive Analysis of Existing Surveys on Security in Drones: Attacks, Countermeasures, and Open Challenges}
\label{Table:Survey_of_existing_survey}
\begin{tabular}{|c|c|c|ccc|c|}
\hline
\multirow{2}{*}{\textbf{Author}}                                        & \multirow{2}{*}{\textbf{Year}} & \multirow{2}{*}{\textbf{Focus of the survey}}                                                                        & \multicolumn{3}{c|}{\textbf{Security}}                                                                                                                                                                                                        & \textbf{Countermeasures}                                         \\ \cline{4-7} 
                                                                        &                                &                                                                                                                      & \multicolumn{1}{c|}{\textbf{Vulnerability}}                                          & \multicolumn{1}{c|}{\textbf{Threat}}                                                 & \textbf{Attack}                                                 &                                                                 \\ \hline
Pham et al.~\cite{pham2015survey}                      & 2015                           & Collision avoidance                                                                                                  & \multicolumn{1}{c|}{\color[HTML]{FE0000} \xmark} & \multicolumn{1}{c|}{\color[HTML]{FE0000} \xmark} & \cmark                                           & \color[HTML]{FE0000} \xmark \\ \hline
Altawy et al.~\cite{altawy2016security}                & 2016                           & Civillian drone security                                                                                             & \multicolumn{1}{c|}{\color[HTML]{FE0000} \xmark} & \multicolumn{1}{c|}{\color[HTML]{FE0000} \xmark} & \cmark                                           & \color[HTML]{FE0000} \xmark \\ \hline
Maxa et al.~\cite{maxa2017survey}                      & 2017                           & Routing Protocols                                                                                                    & \multicolumn{1}{c|}{\color[HTML]{FE0000} \xmark} & \multicolumn{1}{c|}{\cmark}                                           & \color[HTML]{FE0000} \xmark & \color[HTML]{FE0000} \xmark \\ \hline
Lin et al.~\cite{lin2018security}                      & 2018                           & Civillian drone security                                                                                             & \multicolumn{1}{c|}{\color[HTML]{FE0000} \xmark} & \multicolumn{1}{c|}{\color[HTML]{FE0000} \xmark} & \cmark                                           & \color[HTML]{FE0000} \xmark \\ \hline
Fotouhi et al.~\cite{fotouhi2019survey}                & 2019                           & Drone cellular communication                                                                                         & \multicolumn{1}{c|}{\color[HTML]{FE0000} \xmark} & \multicolumn{1}{c|}{\color[HTML]{FE0000} \xmark} & \color[HTML]{FE0000} \xmark & \color[HTML]{FE0000} \xmark \\ \hline
Zhi et al.~\cite{zhi2020security}                      & 2020                           & \begin{tabular}[c]{@{}c@{}}Security and privacy in communication\\ sensors in Multi-UAV setup\end{tabular}           & \multicolumn{1}{c|}{\color[HTML]{FE0000} \xmark} & \multicolumn{1}{c|}{\cmark}                                           & \color[HTML]{FE0000} \xmark & \color[HTML]{FE0000} \xmark \\ \hline
Yaacoub et al.~\cite{yaacoub2020security}              & 2020                           & \begin{tabular}[c]{@{}c@{}}Security and privacy aspects in military \\ and civilian drones\end{tabular}              & \multicolumn{1}{c|}{\color[HTML]{FE0000} \xmark} & \multicolumn{1}{c|}{\color[HTML]{FE0000} \xmark} & \cmark                                           & \color[HTML]{FE0000} \xmark \\ \hline
Nassi et al.~\cite{nassi2021sok}                       & 2021                           & Privacy, security aspects of commercial drones                                                                       & \multicolumn{1}{c|}{\color[HTML]{FE0000} \xmark} & \multicolumn{1}{c|}{\color[HTML]{FE0000} \xmark} & \cmark                                           & \color[HTML]{FE0000} \xmark \\ \hline
Kangunde et al.~\cite{kangunde2021review}              & 2021                           & \begin{tabular}[c]{@{}c@{}}Privacy and security aspects of\\ Real time systems\end{tabular}                          & \multicolumn{1}{c|}{\color[HTML]{FE0000} \xmark} & \multicolumn{1}{c|}{\color[HTML]{FE0000} \xmark} & \color[HTML]{FE0000} \xmark & \color[HTML]{FE0000} \xmark \\ \hline
Boccadoro et al.~\cite{boccadoro2021extensive}         & 2021                           & \begin{tabular}[c]{@{}c@{}}Privacy and security aspects of\\ OSI layer and Cross-layer attacks\end{tabular}          & \multicolumn{1}{c|}{\color[HTML]{FE0000} \xmark} & \multicolumn{1}{c|}{\cmark}                                           & \color[HTML]{FE0000} \xmark & \cmark                                           \\ \hline
Samanth et al.~\cite{samanth2022security}              & 2022                           & Multi-node network IOD security                                                                                      & \multicolumn{1}{c|}{\color[HTML]{FE0000} \xmark} & \multicolumn{1}{c|}{\color[HTML]{FE0000} \xmark} & \color[HTML]{FE0000} \xmark &      \xmark                                                           \\ \hline
Tsao et al.~\cite{tsao2022survey}                      & 2022                           & \begin{tabular}[c]{@{}c@{}}Multi-node network, IOD\\ FANET drone security\end{tabular}                               & \multicolumn{1}{c|}{\color[HTML]{FE0000} \xmark} & \multicolumn{1}{c|}{\color[HTML]{FE0000} \xmark} & \color[HTML]{FE0000} \xmark & \color[HTML]{FE0000} \xmark \\ \hline
Nwaogu et al.~\cite{nwaogu2023application}             & 2023                           & \begin{tabular}[c]{@{}c@{}}Classification of UAV uses in \\ construction, building industry\end{tabular}             & \multicolumn{1}{c|}{\color[HTML]{FE0000} \xmark} & \multicolumn{1}{c|}{\color[HTML]{FE0000} \xmark} & \color[HTML]{FE0000} \xmark & \color[HTML]{FE0000} \xmark \\ \hline
Jones et al.~\cite{jones2023path}                      & 2023                           & Path planning algorithms of UAV                                                                                      & \multicolumn{1}{c|}{\color[HTML]{FE0000} \xmark} & \multicolumn{1}{c|}{\cmark}                                           & \color[HTML]{FE0000} \xmark & \color[HTML]{FE0000} \xmark \\ \hline
Telli et al.~\cite{telli2023comprehensive}             & 2023                           & Aircraft control algorithms                                                                                          & \multicolumn{1}{c|}{\color[HTML]{FE0000} \xmark} & \multicolumn{1}{c|}{\color[HTML]{FE0000} \xmark} & \color[HTML]{FE0000} \xmark & \color[HTML]{FE0000} \xmark \\ \hline
Hafeez et al.~\cite{hafeez2023blockchain}              & 2023                           & Blockchain based UAV Communication                                                                                   & \multicolumn{1}{c|}{\color[HTML]{FE0000} \xmark} & \multicolumn{1}{c|}{\color[HTML]{FE0000} \xmark} & \color[HTML]{FE0000} \xmark & \cmark                                           \\ \hline
Mekdad et al.~\cite{mekdad2023survey}                  & 2023                           & \begin{tabular}[c]{@{}c@{}}Privacy and security in hardware,\\ software, communication and sensor-level\end{tabular} & \multicolumn{1}{c|}{\color[HTML]{FE0000} \xmark} & \multicolumn{1}{c|}{\color[HTML]{FE0000} \xmark} & \cmark                                           & \color[HTML]{FE0000} \xmark \\ \hline
Aretoulaki et al.~\cite{aretoulaki2023complementarity} & 2023                           & Internet-of-drones security                                                                                          & \multicolumn{1}{c|}{\cmark}                                           & \multicolumn{1}{c|}{\color[HTML]{FE0000} \xmark} & \color[HTML]{FE0000} \xmark & \color[HTML]{FE0000} \xmark \\ \hline
Sihag et al.~\cite{sihag2023cyber4drone}               & 2023                           & Drone forensics                                                                                                      & \multicolumn{1}{c|}{\color[HTML]{FE0000} \xmark} & \multicolumn{1}{c|}{\color[HTML]{FE0000} \xmark} & \color[HTML]{FE0000} \xmark & \color[HTML]{FE0000} \xmark \\ \hline
Mykytyn et al.~\cite{mykytyn2024survey}                & 2024                           & Sensor and communication                                                                                             & \multicolumn{1}{c|}{\cmark}                                           & \multicolumn{1}{c|}{\color[HTML]{FE0000} \xmark} & \color[HTML]{FE0000} \xmark & \color[HTML]{FE0000} \xmark \\ \hline
Famili et al.~\cite{famili2024securing}                & 2024                           & Detection Algorithms                                                                                                 & \multicolumn{1}{c|}{\cmark}                                           & \multicolumn{1}{c|}{\color[HTML]{FE0000} \xmark} & \color[HTML]{FE0000} \xmark & \color[HTML]{FE0000} \xmark \\ \hline
Our paper                                                               & 2025                           & Security and privacy in drone systems                                                                                & \multicolumn{1}{c|}{\cmark}                                           & \multicolumn{1}{c|}{\cmark}                                           & \cmark                                           & \cmark                                           \\ \hline
\end{tabular}
\end{table}
\section{Limitations of Existing Literature Surveys} 
\label{sec3_Limitations_of_Existing_Literature_Surveys}
\textbf{Existing Survey Papers on Drone Security:} \label{General Survey}
Nassi et al.~\cite{nassi2021sok} gave an in-depth analysis of the UAV concerning two main questions. First, they discussed whether drones are vulnerable to society or not, and then they also surveyed whether society is a threat to drones or not. They provided a methodical approach with a wide classification of the state-of-the-art prior works. Mekdad et al.~\cite{mekdad2023survey} provided a systematic survey discussing vulnerabilities, attacks, threats, and respective countermeasures for the end-to-end components of a UAV, such as hardware, software, communication link, and sensors. The authors also categorized privacy issues in UAVs for data protection, compliance with drones, policy issues, attacks, and defense methods. Altawy et al.~\cite{altawy2016security} provided a comprehensive review of UAV security, starting from the applications of the UAV, rules, and regulations for managing the airspace, system architecture of drones with a list of various firmware, hardware used, etc., to security and privacy aspects in UAV, protecting confidentiality, integrity, availability, authorization, etc. The challenges regarding privacy preservation, information leakage in drones, and their respective defense mechanism have been discussed. Tu et al.~\cite{tu2024security} studied the security and privacy concerns in the last mile drone delivery services. 

Although the aforementioned surveys provide end-to-end discussions of UAV components such as hardware, software, communication modules, and sensors~\cite{mekdad2023survey, altawy2016security}, they often lack a systematic mapping between specific attacks and the corresponding countermeasures. Others focus narrowly on domains like drone delivery~\cite{tu2024security}, offering depth but lacking a holistic security and privacy perspective.

\noindent
\textbf{Existing Survey Papers on Drone Networks:}
A survey on physical layer security, challenges, issues, threats, and attacks with countermeasures has been employed by Sun et al.~\cite{sun2019physical}. Wang et al.~\cite{wang2021security} investigated three main types of network layer attacks in UAVs that are flooding attacks, routing attacks, and de-authentication attacks. They also discussed two types of physical layer attacks: active eavesdropping attacks and passive eavesdropping attacks, where the detection of the latter is difficult compared to the former. Vasconcelos et al.~\cite{vasconcelos2016impact, vasconcelos2019evaluation} discussed some DDOS attacks in UAV. Salmah et al.~\cite{salamh2019drone} also studied and provided brief evaluation schemes of DoS attacks in UAVs. Skondras et al.~\cite{skondras2021network} proposed a network-slicing framework in UAV networks to secure the UAV networks and to improve high bandwidth and low latency with highly optimized resource allocation. Although certain papers touch on mitigation strategies~\cite{sun2019physical, yang2022review}, few surveys provide a structured or tabular mapping of attacks to countermeasure types, limiting their practical utility for system designers. 

\noindent
\textbf{Survey on IOD, Swarm, FANET:}
The authors~\cite{boccadoro2021extensive} broadly studied Internet-of-drones (IOD) concerning the network architecture, including all possible vulnerabilities in every layer of the Internet protocol stack of the OSI model, starting from the physical layer to the application layer. Cross-layer attacks and optimization approaches for path planning and collision avoidance protocols with UAV security and privacy aspects have been discussed. The authors also study the analysis of the UAVs' economic and social implications, which is novel and gives future research directions.
Yang et al.~\cite{yang2022review} noted security and privacy issues in IOD, and to address countermeasures, they have outlined authentication schemes and blockchain based schemes lightweight authentication, cryptography based authentication schemes, biometric based authentication schemes, etc. Blockchain based schemes are classified for access control, data management, autonomous IOD, etc. 

Tsao et al.~\cite{tsao2022survey} explored various vulnerabilities, threats, and possible solutions with respect to ad-hoc networks, where drones are considered nodes. They carried out a detailed study where the security and privacy of flying ad-hoc networks (FANET) or multi-node networks of drones are considered regarding the communications based on OSI layers. This survey categorizes the attacks and threats based on two types such as nodes, which are drones themselves, and connections of the drones. The paper also provides insights into routing protocols and implementable solutions to the threats concerning security objectives in each layer of the OSI stack. 
\\
Research on multi-layer network attack surfaces is often isolated. Cross-layer perspectives, especially in the context of UAV-specific constraints such as real-time operation, lightweight cryptography, and mobility, are not often unified.

\noindent
\textbf{Survey on Blockchain and SDN:}
Hafeez et al.~\cite{hafeez2023blockchain} detailed all the blockchain-based techniques incorporated for networks of drones in a swarm setting. Renu et al.~\cite {renu2020blockchain} summarized blockchain-based methods deployed in UAV networks. Da et al.~\cite{da2022development} MQTT for control and DDoS detection in UAV systems. Mukherjee et al.~\cite{mukherjee2023isocialdrone} proposed a novel method based on SDN-MQTT-based protocol for UAV systems. Xiong et al.~\cite{xiong2019sdn} studied SDN-based protocols for drone networks. 

\noindent
\textbf{Survey on ROS:}
Recent research has focused on various security aspects in the Robot Operating System (ROS$2$). McClean et al.~\cite{mcclean2013preliminary} conducted a preliminary assessment of vulnerabilities in ROS$2$, highlighting potential security issues that must be addressed. Breiling et al.~\cite{breiling2017secure} proposed a secure communication channel for ROS2 to enhance the protection of data transmitted within the system. Dieber et al.~\cite{dieber2017security} identified several critical security concerns, including unauthorized publishing of data, unauthorized data access, and Denial of Service (DoS) attacks affecting ROS nodes. Diluoffo et al.~\cite{diluoffo2018robot} discussed the trade-offs between performance and security in ROS2, emphasizing the need to balance these two aspects. Deng et al.~\cite{deng2022security} provided an in-depth analysis of security issues in ROS2, contributing to a broader understanding of its vulnerabilities. Lastly, Lee et al.~\cite{lee2021robot} explored secure communication for UAVs, which is relevant for ensuring secure communication in robotic systems, including those utilizing ROS2. Surveys involving emerging technologies like SDN~\cite{mukherjee2023isocialdrone}, blockchain~\cite{hafeez2023blockchain}, and ROS~\cite{deng2022security} present promising advancements but are still in early stages of exploring attack resilience mechanisms and deployment challenges in real world UAV settings. The above discussion and analysis highlights the need for a comprehensive, component wise systematization of attacks and countermeasures. Our work addresses this by organizing UAV threats by component (hardware, software, communication, sensors/actuators) and mapping them to well categorized countermeasure groups in a many to many fashion.

\noindent
\textbf{Summary:} Existing surveys provide valuable insights into UAV security, yet they remain fragmented in scope and depth. Some works classify attacks across multiple UAV components but fail to consolidate or distill actionable insights, while others focus narrowly on domains such as drone delivery or flying ad-hoc networks, lacking a holistic system-wide perspective. Moreover, few studies explicitly map attacks to countermeasures, leaving the effectiveness of defenses unclear. To address this, our work presents a comprehensive taxonomy of 40 UAV-specific attacks mapped to 43 countermeasures, enabling clear traceability between threats and their mitigation. 

To address these gaps, this work is guided by three research questions:
\begin{itemize}
    \item \textbf{RQ1:} What types of attacks target UAV systems across different layers, and how do they impact operations, security, and regulatory compliance?
    \item \textbf{RQ2:} What countermeasures exist to protect UAVs, and do they address the existing attacks?
    \item \textbf{RQ3:} How can a regulatory authority securely monitor UAVs, detect malicious behavior, and enforce policy-compliant commands in real time?
\end{itemize}


\section{Systematization of Attacks}\label{sec4_Systematization_of_Attacks}
UAV systems are vulnerable to several attacks ranging from the physical layer to the higher layers of the network protocol stack. For instance, in GPS spoofing, attackers transmit false GPS signals to mislead the drone's navigation. GPS jamming disrupts GPS signals, preventing accurate positioning. Communication interception involves capturing and potentially altering data between the UAV and its operator. Command injection allows unauthorized commands to control the UAV. Malware can corrupt the UAV’s software, affecting its operation or data security. Denial of Service (DoS) attacks overload communication systems or processors to incapacitate the UAV.

\subsection{Methodology}
Unmanned Aerial Vehicles (UAVs) are susceptible to various cyber and physical attacks. These attacks can be assessed based on multiple parameters, including 
\textit{impact (confidentiality, integrity, and availability)},
\textit{severity},
\textit{mission impact}, \textit{irrecoverability}, \textit{stealthiness}, and \textit{exploitation complexity}. This section thoroughly explores these properties and provides a framework for evaluating UAV-related attacks.
\subsubsection{Impact (Confidentiality/ Integrity/ Availability)}
Impact refers to the security impact that the attack exploits. It evaluates the extent to which an attack affects the system’s ability to perform its intended tasks. The impact is categorized into three key aspects such as, {\em Confidentiality (C)} ensures that sensitive UAV data is accessible only to authorized entities, preventing unauthorized access through encryption and access control mechanisms. {\em Integrity (I)} guarantees that UAV data and communications remain unaltered during transmission and storage, protecting against tampering, injection, or corruption using cryptographic techniques and verification mechanisms. {\em Availability (A)} ensures the continuous operation of UAV services, communication links, and control mechanisms by defending against disruptions such as denial-of-service (DoS) attacks, jamming, and other forms of interference.
\\
\subsubsection{Severity}
Severity is a critical metric that quantifies the extent of damage caused by an attack. It measures the impact on UAV operations, safety, and functionality.
\begin{itemize}[noitemsep]
\item \textbf{High~(\fullcirc)}: High-severity attacks cause significant damage, potentially leading to complete loss of control, mission failure, or severe security breaches.
\item \textbf{Medium~(\halfcirc)}: Medium-severity attacks disrupt UAV operations but do not cause total failure. The system may recover with some effort, though performance and security may be compromised.
\item \textbf{Low~(\emptycirc)}: Low-severity attacks have minimal impact, causing minor disruptions or data exposure without critically affecting UAV functionality.
\end{itemize}
\subsubsection{Mission Disruption}
Mission disruption refers to the extent to which an attack interrupts or prevents the UAV from completing its intended mission.
\begin{itemize}[noitemsep]
    \item \textbf{High~(\fullcirc)}: A high mission impact attack results in the UAV losing control, crashing, or becoming entirely unable to complete its mission.
    \item \textbf{Medium~(\halfcirc)}: Medium mission impact attacks hinder UAV operations but do not entirely prevent mission completion. The UAV may experience malfunctions or temporary disruptions. 
    \item \textbf{Low~(\emptycirc)}: Low mission impact attacks do not directly affect mission completion. The UAV remains operational while they may lead to data theft or minor disruptions.
\end{itemize}

\subsubsection{Irrecoverability}
Irrecoverability measures the difficulty of restoring the UAV system to a secure and operational state after an attack.

\begin{itemize}[noitemsep]
    \item \textbf{High~(\fullcirc)}: A highly irrecoverable attack causes permanent damage, requiring hardware replacement, system reconfiguration, or complete restoration. 
    \item \textbf{Medium~(\halfcirc)}: Medium irrecoverability attacks cause significant disruptions but allow recovery through backups or security patches.
    \item \textbf{Low~(\emptycirc)}: Low irrecoverability attacks cause minimal long-term damage. The UAV can recover quickly with minor interventions.
\end{itemize}

\subsubsection{Stealthiness}
Stealthiness refers to how easily an attack can evade detection by UAV security systems.

\begin{itemize}[noitemsep]
    \item \textbf{High~(\fullcirc)}: Highly stealthy attacks are difficult to detect and often bypass security mechanisms. They avoid triggering alarms and operate covertly. 
    \item \textbf{Medium~(\halfcirc)}: Medium stealth attacks may evade some security measures but leave detectable traces. Careful analysis can uncover their presence.
    \item \textbf{Low~(\emptycirc)}: Low stealth attacks generate noticeable disruptions, triggering security alarms. These attacks are relatively easy to detect. 
\end{itemize}

\subsubsection{Exploitation Complexity}
Exploitation complexity measures the level of expertise, resources, and sophistication required to execute an attack.

\begin{itemize}[noitemsep]
    \item \textbf{High~(\fullcirc)}: High complexity attacks involve multiple layers of exploitation, requiring advanced knowledge, specialized tools, and significant reconnaissance.
    \item \textbf{Medium~(\halfcirc)}: Medium complexity attacks require technical knowledge but are not as intricate as high-complexity attacks. They may use advanced tools but do not require multi-layered exploitation. 
    \item \textbf{Low~(\emptycirc)}: Low complexity attacks are relatively simple to execute, often targeting well-known vulnerabilities. They require minimal expertise and resources.
\end{itemize}
\par
An attack on a UAV can be evaluated based on the key parameters discussed above. In general, high mission impact attacks cause severe operational failures, while low impact attacks primarily affect data integrity. Attacks with high irrecoverability require extensive recovery efforts, whereas low irrecoverability attacks allow for quick restoration. Highly stealthy attacks evade detection, while low stealth attacks are easily noticed. Lastly, high-complexity attacks demand significant expertise, while low-complexity attacks exploit known vulnerabilities with minimal effort. By analyzing these parameters in Table~\ref{Table:Attack}, UAV operators can classify and prioritize threats, enabling effective countermeasures against potential cyber attacks on UAV systems.
\begin{sidewaystable*}[htbp]
\scriptsize
\caption{Attack classification: This table categorizes various attacks targeting UAV systems based on their attack target, type, method, target, and potential damage. It also highlights the impact on confidentiality (C), integrity (I), and availability (A), along with associated metrics such as mission disruption, irrecoverability, stealthiness, and exploitation complexity.}
\label{Table:Attack}
\begin{tabular}{|l|l|l|l|ccc|c|c|c|c|c|}
\hline
\multirow{2}{*}{\textbf{Targets}}                                                                       & \multirow{2}{*}{\textbf{Type}}                                                                                                  & \multirow{2}{*}{\textbf{Vulnerability}} & \multirow{2}{*}{\textbf{Damage}} & \multicolumn{3}{c|}{\textbf{Impact}}                                                                                                 & \multirow{2}{*}{\textbf{Severity}} & \multirow{2}{*}{\textbf{\begin{tabular}[c]{@{}c@{}}Mission\\ Impact\end{tabular}}} & \multirow{2}{*}{\textbf{Irrecoverability}} & \multirow{2}{*}{\textbf{Stealthiness}} & \multirow{2}{*}{\textbf{\begin{tabular}[c]{@{}c@{}}Exploitation\\ Complexity\end{tabular}}} \\ \cline{5-7}
                                                                                                        &                                                                                                                                 &                                         &                                  & \multicolumn{1}{c|}{\textbf{C}}            & \multicolumn{1}{c|}{\textbf{I}}            & \textbf{A}                                 &                                    &                                                                                    &                                            &                                        &                                                                                             \\ \hline
\multirow{7}{*}{\textbf{\begin{tabular}[c]{@{}l@{}}T1: Wireless\\ Communication\\ Module\end{tabular}}} & A1: Passive Eavesdropping~\cite{sun2019physical}                                                          & Plaintext transmissions                 & Data Interception                & \multicolumn{1}{l|}\cmark & \multicolumn{1}{l|}{\color[HTML]{FE0000} \xmark} & \multicolumn{1}{l|}{\color[HTML]{FE0000} \xmark} & \emptycirc                                  & \emptycirc                                                                                  & \emptycirc                                          & \fullcirc                                      & \emptycirc                                                                                           \\ \cline{2-12} 
                                                                                                        & A2: Active Eavesdropping~\cite{lu2018proactive}                                                           & Unsecured comm. channel         & Data Theft                       & \multicolumn{1}{l|}\cmark & \multicolumn{1}{l|}\cmark & \multicolumn{1}{l|}\cmark & \halfcirc                                  & \halfcirc                                                                                  & \emptycirc                                          & \halfcirc                                      & \emptycirc                                                                                           \\ \cline{2-12} 
                                                                                                        & A3: Signal Interception~\cite{abughalwa2020full}                                                          & Weak encryption                         & Data Theft                       & \multicolumn{1}{l|}\cmark & \multicolumn{1}{l|}\cmark & \multicolumn{1}{l|}\cmark & \halfcirc                                  & \halfcirc                                                                                  & \emptycirc                                          & \halfcirc                                      & \halfcirc                                                                                           \\ \cline{2-12} 
                                                                                                        & A4: Wi-Fi Deauthentication~\cite{kadripathi2020authentication}                                                                                                      & Weak Wi-Fi authentication               & Data Theft                       & \multicolumn{1}{l|}{\color[HTML]{FE0000} \xmark} & \multicolumn{1}{l|}{\color[HTML]{FE0000} \xmark} & \multicolumn{1}{l|}\cmark & \fullcirc                                  & \fullcirc                                                                                  & \emptycirc                                          & \emptycirc                                      & \emptycirc                                                                                           \\ \cline{2-12} 
                                                                                                        & A5: Bluetooth Exploits~\cite{bluetooth_exploits}                                                                                                          & Weak Bluetooth authentication           & Data Interception                & \multicolumn{1}{l|}\cmark & \multicolumn{1}{l|}\cmark & \multicolumn{1}{l|}\cmark & \fullcirc                                  & \fullcirc                                                                                  & \halfcirc                                          & \halfcirc                                      & \halfcirc                                                                                           \\ \cline{2-12} 
                                                                                                        & A6: False MAVLink Injection~\cite{jeong2021muvids}                                                                                                     & Unauthenticated MAVLink        & System Manipulation              & \multicolumn{1}{l|}{\color[HTML]{FE0000} \xmark} & \multicolumn{1}{l|}\cmark & \multicolumn{1}{l|}\cmark & \fullcirc                                  & \fullcirc                                                                                  & \halfcirc                                          & \halfcirc                                      & \halfcirc                                                                                           \\ \cline{2-12} 
                                                                                                        & A7:  Radio-Frequency Jamming~\cite{mekdad2024exploring}                                                                                                    & Unsecured RF channels                   & Service Disruption               & \multicolumn{1}{l|}{\color[HTML]{FE0000} \xmark} & \multicolumn{1}{l|}{\color[HTML]{FE0000} \xmark} & \multicolumn{1}{l|}\cmark & \fullcirc                                  & \fullcirc                                                                                  & \emptycirc                                          & \emptycirc                                      & \emptycirc                                                                                           \\ \hline
\multirow{6}{*}{\textbf{T2: Sensors}}                                                                   & A8: GPS Spoofing~\cite{xue2020deepsim}                                                                    & GNSS authentication                     & Navigation Manipulation          & \multicolumn{1}{c|}{\color[HTML]{FE0000} \xmark} & \multicolumn{1}{c|}\cmark & \cmark                      & \fullcirc                                  & \fullcirc                                                                                  & \halfcirc                                          & \fullcirc                                      & \halfcirc                                                                                           \\ \cline{2-12} 
                                                                                                        & A9: GPS Jamming~\cite{mekdad2024exploring}                                                                & Low-power GNSS signals                  & Navigation Disruption            & \multicolumn{1}{c|}{\color[HTML]{FE0000} \xmark} & \multicolumn{1}{c|}{\color[HTML]{FE0000} \xmark} & \cmark                      & \fullcirc                                  & \fullcirc                                                                                  & \emptycirc                                          & \emptycirc                                      & \emptycirc                                                                                           \\ \cline{2-12} 
                                                                                                        & A10: Camera Spoofing~\cite{nassi2019mobilbye}                                                             & Outdated firmware                       & False Image Injection            & \multicolumn{1}{c|}{\color[HTML]{FE0000} \xmark} & \multicolumn{1}{c|}\cmark & \color[HTML]{FE0000} \xmark                      & \halfcirc                                  & \halfcirc                                                                                  & \emptycirc                                          & \fullcirc                                      & \halfcirc                                                                                           \\ \cline{2-12} 
                                                                                                        & A11: Acoustic Injection~\cite{jeong2023rocking}                                                           & Vulnerable MEMS sensors                 & Sensor Malfunction               & \multicolumn{1}{c|}{\color[HTML]{FE0000} \xmark} & \multicolumn{1}{c|}\cmark & \cmark                      & \halfcirc                                  & \halfcirc                                                                                  & \emptycirc                                          & \fullcirc                                      & \halfcirc                                                                                           \\ \cline{2-12} 
                                                                                                        & A12: LiDAR Spoofing~\cite{khan2024lidar}                                                                                                             & Reliance on raw sensors input           & Navigation Manipulation          & \multicolumn{1}{c|}{\color[HTML]{FE0000} \xmark} & \multicolumn{1}{c|}\cmark & \color[HTML]{FE0000} \xmark                      & \halfcirc                                  & \halfcirc                                                                                  & \emptycirc                                          & \fullcirc                                      & \halfcirc                                                                                           \\ \cline{2-12} 
                                                                                                        & A13: Sensor Deprivation Attack~\cite{erba2024sensor}                                                                                                  & Single Sensor Dependency        & Sensor Malfunction               & \multicolumn{1}{c|}{\color[HTML]{FE0000} \xmark} & \multicolumn{1}{c|}{\color[HTML]{FE0000} \xmark} & \cmark                      & \fullcirc                                  & \fullcirc                                                                                  & \halfcirc                                          & \halfcirc                                      & \halfcirc                                                                                           \\ \hline
\multirow{8}{*}{\textbf{\begin{tabular}[c]{@{}l@{}}T3: Hardware\\ and \\Firmware\end{tabular}}}           & \begin{tabular}[c]{@{}l@{}}A14: Electro-Magnetic Injection~\cite{jang2023paralyzing}\end{tabular} & Lack of hardware shielding              & System Malfunction               & \multicolumn{1}{c|}{\color[HTML]{FE0000} \xmark} & \multicolumn{1}{c|}\cmark & \cmark                      & \fullcirc                                  & \fullcirc                                                                                  & \halfcirc                                          & \halfcirc                                      & \fullcirc                                                                                           \\ \cline{2-12} 
                                                                                                        & A15: Side-Channel Attack~\cite{radtke_host_2022}                                                                                                        & Information leakage                     & System Data Theft                & \multicolumn{1}{c|}\cmark & \multicolumn{1}{c|}{\color[HTML]{FE0000} \xmark} & \color[HTML]{FE0000} \xmark                      & \fullcirc                                  & \halfcirc                                                                                  & \fullcirc                                          & \fullcirc                                      & \fullcirc                                                                                           \\ \cline{2-12} 
                                                                                                        & A16: Firmware Hacking~\cite{kim2024challenges}                                                            & Insecure firmware                       & System Compromise                & \multicolumn{1}{c|}{\color[HTML]{FE0000} \xmark} & \multicolumn{1}{c|}\cmark & \cmark                      & \fullcirc                                  & \fullcirc                                                                                  & \fullcirc                                          & \halfcirc                                      & \fullcirc                                                                                           \\ \cline{2-12} 
                                                                                                        & A17: Hardware Trojan~\cite{mynuddin2024trojan}                                                                                                            & Untrusted supply chain                  & System Malfunction               & \multicolumn{1}{c|}\cmark & \multicolumn{1}{c|}\cmark & \cmark                      & \fullcirc                                  & \fullcirc                                                                                  & \fullcirc                                          & \fullcirc                                      & \fullcirc                                                                                           \\ \cline{2-12} 
                                                                                                        & A18: Bootloader Exploit~\cite{schiller2023drone}                                                                                                         & No secure boot                          & System Compromise                & \multicolumn{1}{c|}{\color[HTML]{FE0000} \xmark} & \multicolumn{1}{c|}\cmark & \cmark                      & \fullcirc                                  & \fullcirc                                                                                  & \fullcirc                                          & \halfcirc                                      & \fullcirc                                                                                           \\ \cline{2-12} 
                                                                                                        & A19: Battery Depletion~\cite{desnitsky2021simulation}                                                                                                         & Weak battery protection                 & Power Dissrupption               & \multicolumn{1}{c|}{\color[HTML]{FE0000} \xmark} & \multicolumn{1}{c|}{\color[HTML]{FE0000} \xmark} & \cmark                      & \fullcirc                                  & \fullcirc                                                                                  & \halfcirc                                          & \emptycirc                                      & \halfcirc                                                                                           \\ \cline{2-12} 
                                                                                                        & A20: Fault Injection~\cite{ioactiveDroneSecurity}                                                                                                            & Physical access                         & System Malfunction               & \multicolumn{1}{l|}{\color[HTML]{FE0000} \xmark} & \multicolumn{1}{l|}\cmark & \multicolumn{1}{l|}\cmark & \fullcirc                                  & \fullcirc                                                                                  & \fullcirc                                          & \halfcirc                                      & \fullcirc                                                                                           \\ \cline{2-12} 
                                                                                                        & A21: Row-Hammer Attack                                                                                                          & DRAM flaws                              & Data corruption                  & \multicolumn{1}{l|}\cmark & \multicolumn{1}{l|}\cmark & \multicolumn{1}{l|}{\color[HTML]{FE0000} \xmark} & \fullcirc                                  & \fullcirc                                                                                  & \halfcirc                                          & \fullcirc                                      & \halfcirc                                                                                           \\ \hline
\multirow{10}{*}{\textbf{\begin{tabular}[c]{@{}l@{}}T4: Operating \\ Systems and\\ Application\end{tabular}}}       & A22: Malware~\cite{jares2021investigating}                                                                & Software integrity                      & Unauthorized Access              & \multicolumn{1}{c|}\cmark & \multicolumn{1}{c|}\cmark & \cmark                      & \halfcirc                                  & \fullcirc                                                                                  & \fullcirc                                          & \fullcirc                                      & \halfcirc                                                                                           \\ \cline{2-12} 
                                                                                                        & A23: DL-based Trojans~\cite{mynuddin2024trojan}                                                           & DL Model poisoning                      & Unauthorized Control             & \multicolumn{1}{c|}{\color[HTML]{FE0000} \xmark} & \multicolumn{1}{c|}\cmark & \color[HTML]{FE0000} \xmark                      & \fullcirc                                  & \fullcirc                                                                                  & \fullcirc                                          & \fullcirc                                      & \fullcirc                                                                                    \\ \cline{2-12} 
                                                                                                        & A24: Code Injection~\cite{habibi2015mavr}                                                                 & Input validation                        & System Compromise                & \multicolumn{1}{c|}{\color[HTML]{FE0000} \xmark} & \multicolumn{1}{c|}\cmark & \cmark                      & \fullcirc                                  & \fullcirc                                                                                  & \halfcirc                                          & \halfcirc                                      & \halfcirc                                                                                           \\ \cline{2-12} 
                                                                                                        & A25: Code Reuse~\cite{habibi2015mavr}                                                                                                                 & Lack of control flow integrity          & System Compromise                & \multicolumn{1}{c|}{\color[HTML]{FE0000} \xmark} & \multicolumn{1}{c|}\cmark & \cmark                      & \fullcirc                                  & \halfcirc                                                                                  & \emptycirc                                          & \halfcirc                                      & \emptycirc                                                                                           \\ \cline{2-12} 
                                                                                                        & A26: Buffer Overflow~\cite{snykSnykVulnerability}                                                                                                            & Poor bounds checking                    & System Compromise                & \multicolumn{1}{c|}{\color[HTML]{FE0000} \xmark} & \multicolumn{1}{c|}\cmark & \cmark                      & \halfcirc                                  & \fullcirc                                                                                  & \halfcirc                                          & \emptycirc                                      & \halfcirc                                                                                           \\ \cline{2-12} 
                                                                                                        & A27: Zero-Day Attack~\cite{orhun2023hybrid}                                                                                                            & unpatched software flaws      & System Malfunction               & \multicolumn{1}{c|}\cmark & \multicolumn{1}{c|}\cmark & \cmark                      & \fullcirc                                  & \fullcirc                                                                                  & \fullcirc                                          & \fullcirc                                      & \fullcirc                                                                                           \\ \cline{2-12} 
                                                                                                        & A28: Data Poisoning                                                                                                             & Lack of ML data validation           & Data MAnipulation                & \multicolumn{1}{c|}{\color[HTML]{FE0000} \xmark} & \multicolumn{1}{c|}\cmark & \color[HTML]{FE0000} \xmark                      & \fullcirc                                  & \fullcirc                                                                                  & \fullcirc                                          & \fullcirc                                      & \fullcirc                                                                                           \\ \cline{2-12} 
                                                                                                        & A29: Adversarial Attacks~\cite{tian2021adversarial}                                                       & ML model overfitting                    & Data Manipulation                & \multicolumn{1}{c|}{\color[HTML]{FE0000} \xmark} & \multicolumn{1}{c|}\cmark & \color[HTML]{FE0000} \xmark                      & \fullcirc                                  & \halfcirc                                                                                  & \emptycirc                                          & \fullcirc                                      & \halfcirc                                                                                           \\ \cline{2-12} 
                                                                                                        & A30: Evasion Attacks                                                                                                            & Weak Feature Selection      & Data Manipulation                & \multicolumn{1}{c|}{\color[HTML]{FE0000} \xmark} & \multicolumn{1}{c|}\cmark & \color[HTML]{FE0000} \xmark                      & \halfcirc                                  & \halfcirc                                                                                  & \emptycirc                                          & \fullcirc                                      & \halfcirc                                                                                           \\ \cline{2-12} 
                                                                                                        & A31: Backdoor Attacks~\cite{islam2022triggerless}                                                                                                           & Hidden code paths                       & Data Theft                       & \multicolumn{1}{c|}\cmark & \multicolumn{1}{c|}\cmark & \cmark                      & \halfcirc                                  & \fullcirc                                                                                  & \fullcirc                                          & \fullcirc                                      & \fullcirc                                                                                           \\ \hline
\multirow{9}{*}{\textbf{T5: Networks}}                                                                  & A32: MAC Spoofing~\cite{watkins2018exploiting}                                                            & Identity spoofing                       & Impersonation                    & \multicolumn{1}{c|}{\color[HTML]{FE0000} \xmark} & \multicolumn{1}{c|}\cmark & \color[HTML]{FE0000} \xmark                      & \fullcirc                                  & \halfcirc                                                                                  & \emptycirc                                          & \halfcirc                                      & \emptycirc                                                                                           \\ \cline{2-12} 
                                                                                                        & A33: Packet Sniffing~\cite{westerlund2019drone}                                                           & Unencrypted communication               & Data Theft                       & \multicolumn{1}{c|}\cmark & \multicolumn{1}{c|}{\color[HTML]{FE0000} \xmark} & \color[HTML]{FE0000} \xmark                      & \halfcirc                                  & \halfcirc                                                                                  & \emptycirc                                          & \fullcirc                                      & \emptycirc                                                                                           \\ \cline{2-12} 
                                                                                                        & A34: DoS~\cite{feng2021denial}                                                                            & Lack of TCP state control               & Service Disruption               & \multicolumn{1}{c|}{\color[HTML]{FE0000} \xmark} & \multicolumn{1}{c|}{\color[HTML]{FE0000} \xmark} & \cmark                      & \fullcirc                                  & \fullcirc                                                                                  & \emptycirc                                          & \emptycirc                                      & \emptycirc                                                                                           \\ \cline{2-12} 
                                                                                                        & A35: Black Hole~\cite{xiong2023sbha}                                                                      & Trust-based routing protocols           & Packet Drop                      & \multicolumn{1}{l|}{\color[HTML]{FE0000} \xmark} & \multicolumn{1}{l|}\cmark & \multicolumn{1}{l|}\cmark & \fullcirc                                  & \fullcirc                                                                                  & \halfcirc                                          & \halfcirc                                      & \halfcirc                                                                                           \\ \cline{2-12} 
                                                                                                        & A36: Gray Hole~\cite{ceviz2021analysis}                                                                   & multi-hop routing protocols             & Partial Packet Drop              & \multicolumn{1}{l|}{\color[HTML]{FE0000} \xmark} & \multicolumn{1}{l|}\cmark & \multicolumn{1}{l|}\cmark & \halfcirc                                  & \halfcirc                                                                                  & \halfcirc                                          & \halfcirc                                      & \halfcirc                                                                                           \\ \cline{2-12} 
                                                                                                        & A37: Sybil Attack~\cite{bhandarkar2022adversarial}                                                        & Lack of identity checks                 & False Identity Injection         & \multicolumn{1}{l|}{\color[HTML]{FE0000} \xmark} & \multicolumn{1}{l|}\cmark & \multicolumn{1}{l|}\cmark & \fullcirc                                  & \fullcirc                                                                                  & \halfcirc                                          & \fullcirc                                      & \halfcirc                                                                                           \\ \cline{2-12} 
                                                                                                        & A38: MITM~\cite{li2020lightweight}                                                                        & Authentication                          & Data Eavesdropping               & \multicolumn{1}{l|}\cmark & \multicolumn{1}{l|}\cmark & \multicolumn{1}{l|}\cmark & \fullcirc                                  & \fullcirc                                                                                  & \halfcirc                                          & \fullcirc                                      & \halfcirc                                                                                           \\ \cline{2-12} 
                                                                                                        & A39: Replay Attack~\cite{omar2024sdr}                                                                     & Session freshness                       & Unauthorized Data Reuse       & \multicolumn{1}{l|}{\color[HTML]{FE0000} \xmark} & \multicolumn{1}{l|}\xmark & \multicolumn{1}{l|}\xmark & \halfcirc                                  & \halfcirc                                                                                  & \emptycirc                                          & \halfcirc                                      & \emptycirc                                                                                           \\ \cline{2-12} 
                                                                                                        & A40: DDoS~\cite{clarke2023distributed}                                                                    & Traffic filtering                       & System Overload                  & \multicolumn{1}{l|}{\color[HTML]{FE0000} \xmark} & \multicolumn{1}{l|}{\color[HTML]{FE0000} \xmark} & \multicolumn{1}{l|}\xmark & \fullcirc                                  & \fullcirc                                                                                  & \emptycirc                                          & \emptycirc                                      & \emptycirc                                                                                           \\ \hline
\end{tabular}
\begin{tablenotes}
      \item \text{Note: Where, C = Confidentiality, I = Integrity, A = Availability, \cmark = Satisfactory, {\color[HTML]{FE0000} \xmark} = Unsatisfactory, \fullcirc~= High, \halfcirc~= Medium, \emptycirc~= Low, BLE= Bluetooth Low Energy, MITM = Man-in-the-middle}
      \item \text{ }
\end{tablenotes}
\end{sidewaystable*}
\subsection{Discussion on Critical UAV Attacks}
The following section discusses significant UAV attacks, analyzing their criticality, execution techniques, and impact on UAV security, privacy, and overall system integrity. Understanding these attacks helps in identifying vulnerabilities and developing countermeasures.
\subsubsection{Attacks on Hardware and Software:}
\noindent
\textbf{Malware and Keylogger Attack: }Malware and keyloggers compromise UAV system integrity by stealing operator credentials or corrupting software \cite{rahman2022detection}. Countermeasures include regular firmware updates, anti-malware software, network segmentation, and application whitelisting to prevent unauthorized execution.
\\
\noindent
\textbf{Code Injection Attack: } A Code Injection Attack~\cite{habibi2015mavr} occurs when an attacker exploits vulnerabilities in a drone's software to insert and execute malicious code. This injected code is referred to as malicious application, which can lead to unauthorized command execution, system crashes, or complete control over the drone.  
\\
\noindent
\textbf{Side-channel Attack: }Side-channel attacks~\cite{radtke2022safeguarding} exploit UAV power consumption, electromagnetic emissions, and timing information to extract sensitive data. Defense strategies involve masking techniques, noise generation, regular key updates, and continuous monitoring for anomalies to prevent unauthorized information leakage.
\\
\noindent
\textbf{Deep Learning (DL)-based Trojan Attack: }
DL-based trojan attack~\cite{mynuddin2024trojan} embeds hidden malicious behavior into a deep learning model used in drones for navigation and decision-making. Attackers manipulate the training process by injecting backdoors that trigger specific actions under predefined conditions.
\\
\noindent
\textbf{Adversarial Attack: }Adversarial attacks~\cite{bhandarkar2022adversarial} exploit machine learning models used in drone systems by introducing carefully crafted perturbations to deceive the model. These attacks can cause misclassifications in object detection, navigation errors, or security breaches in AI-driven UAV applications. Zhang et al. ~\cite{zhang2022adversarial} proposed an optimized adversarial patch attack, which can affect multi-scale object detection in UAV remote sensing showing the scalability of the attack is very much high.

\subsubsection{Attacks on Networks}
\noindent
\textbf{Remote Hijacking Attack: }
Remote hijacking occurs when attackers exploit communication protocol flaws, weak authentication, or software vulnerabilities to gain unauthorized control over UAVs \cite{pratama2023behind, dash2021pid}. Countermeasures include encryption, strong authentication, intrusion detection, and signal jamming detection systems. Regular software updates and operator awareness are crucial to mitigate risks.
\\
\noindent
\textbf{Denial-of-Service (DoS)/ Distributed Denial-of-Service (DDoS) Attack: }DoS and DDoS attacks flood UAV control networks with excessive traffic, causing service disruptions \cite{rahman2022detection}. UAVs can become unresponsive, compromising missions. Defense strategies include traffic filtering, redundant communication channels, and anomaly detection. Further analysis is available in \cite{adedeji2023ddos}.
\\
\noindent
\textbf{Cross-layer Attack: }These attacks exploit vulnerabilities across multiple network layers, affecting confidentiality, availability, and integrity \cite{zhai2023hotd}. Defense mechanisms include strong authentication, encryption, intrusion detection, and behavioral anomaly monitoring. Zhai et al. proposed a detection method using supervised learning.
\\
\noindent
\textbf{Replay Attack: }Replay attacks involve intercepting and retransmitting UAV control commands, leading to unauthorized drone manipulation. This can cause mission deviation or malicious activity. Preventive measures include cryptographic timestamps, secure authentication, and sequence number validation.
\\
\noindent
\textbf{Man-in-the-Middle Attack: }MITM attacks occur when an adversary intercepts and alters UAV communications \cite{wlazlo2021man}. Li et al. \cite{li2020lightweight} proposed a lightweight digital signature-based solution to secure drone-GCS communication against MITM threats.

\subsubsection{Attacks on Wireless Communication Module}
\noindent
\textbf{Eavesdropping Attack: }Unauthorized interception of UAV communications can compromise privacy and mission integrity, exposing sensitive flight data \cite{abughalwa2019comparative}. This can be exploited for espionage, sabotage, or unauthorized access. Hoang et al. \cite{hoang2019detection} proposed a detection technique to mitigate eavesdropping attacks.
\\
\noindent
\textbf{Jamming Attack: }Jamming attacks disrupt UAV navigation by overpowering GPS signals with strong radio interference \cite{van2018keeping}. This can affect civilian and military operations. Anti-jamming techniques such as frequency hopping, adaptive filtering, and redundant navigation systems enhance resilience.

\subsubsection{Attacks on Sensors}
\textbf{Spoofing Attack: }GPS spoofing misleads UAV navigation by feeding false positioning signals, leading to mission failure or security breaches \cite{eldosouky2019drones, saputro2020implementation, yao2023swarmfuzz}. Protection measures include cryptographic authentication, multi-sensor fusion for location verification, and anomaly detection mechanisms.
\\
\noindent
\textbf{Acoustic Injection: }Acoustic Injection attacks exploit the UAV sensors by using sound waves. Jeong et al.~\cite{jeong2023rocking} proposed an acoustic injection attack on the UAV sensors and a mitigation technique.

\section{Systematization of Countermeasures} \label{sec5_Systematization_of_Countermeasures}
To protect UAV systems from security threats, various countermeasures can be adopted: (1) implement data encryption and authentication for all communication channels to safeguard against unauthorized access and command manipulation; (2) regularly update firmware and security patches to fix vulnerabilities and prevent malware; (3) ensure physical security for the UAV and encrypt data at rest to enhance overall safety; (4) use anti-spoofing technologies to verify GPS signal integrity and employ jamming detection systems to counteract signal interference. These integrated measures provide a comprehensive security approach for UAV systems. A systematic examination of these countermeasures is vital for evaluating their effectiveness in protecting against emerging threats. Additionally, continuous monitoring and assessment of the security landscape are essential to adapt to new vulnerabilities and threats. Table~\ref{Table:Countermeasures} presents a systematization of the countermeasures with the following described methodology.

\subsection{Methodology}
 To review the countermeasure schemes, we have adopted six specific parameters. The assessment methods are briefly elaborated below. The evaluation of these parameters has been done as high (\fullcirc), medium (\halfcirc), and low (\emptycirc), providing a clear framework for understanding their relative effectiveness.  
 
\subsubsection{Model (Prediction/ Detection/ Response)}
According to the NIST Cyber Security Framework~\cite{NISTpascoe2023public}, we have organized the countermeasures as categorized as Prediction, Detection, and Response. In our work, we have organized the countermeasures with the following criteria. \textbf{Prediction} methods focus on proactively preventing attacks before they occur through mechanisms. \textbf{Detection} methods entail identifying security breaches in real-time or post-event. \textbf{Response} methods define the actions taken upon detecting an attack to mitigate its impact, which may include isolating compromised components, or initiating recovery mechanisms to restore system integrity and functionality.

\subsubsection{Implementation Overhead}
Implementation overhead evaluates the resource consumption and performance impact of security mechanisms on UAV systems. It can be categorized as computation overhead, memory overhead, communication overhead, and resulting latency overhead. \textbf{Computational Overhead} is defined by the costs required to deploy a security measure. Includes CPU, memory overhead, size of the secret key, etc. for the deployment of the countermeasure methods. \textbf{Memory Overhead} is additional storage space required for security mechanisms. It includes encryption keys, logs, and security policies. \textbf{Communication Overhead} refers to the extra network traffic required to send, receive, and process the information effectively. It is typically calculated in terms of the size of the key required for the encryption or the authentication mechanism used in the countermeasure scheme. Finally, the \textbf{Latency Overhead} is described by the delay introduced by a security mechanism in processing or communication. High latency can affect UAV real-time responsiveness.
\begin{itemize}[noitemsep]
    \item \textbf{High~(\fullcirc)} stands for the mechanisms with high computational, memory, latency, and communication overhead significantly impact UAV performance. These mechanisms require extensive CPU resources, large memory storage, high processing time, and increased network bandwidth.
    \item \textbf{Medium~(\halfcirc)} refers to the medium overhead security mechanisms moderately affect performance by requiring some additional computational power and memory but do not severely impact UAV functionality. 
    \item \textbf{Low~(\emptycirc)} justifies the mechanisms with low overhead introduce minimal computational, memory, and communication burden on UAVs. Safeguard methods are expected to be designed for efficiency with negligible performance impact.
\end{itemize}

\subsubsection{Deployability}
Deployability measures how easily a security mechanism can be integrated into existing UAV systems. 
\begin{itemize}[noitemsep]
    \item \textbf{High~(\fullcirc)} deployability stands for the security mechanisms require minimal changes to UAV software or hardware, allowing easy integration. 
    \item \textbf{Medium~(\halfcirc)} deployability refers to the mechanisms may require firmware updates or minor modifications to UAV architecture but do not demand extensive redesign.
    \item \textbf{Low~(\emptycirc)} deployability defines the mechanisms with low deployability require significant hardware modifications or a complete system overhaul. These methods may not be feasible for existing UAV systems.
\end{itemize}
\subsubsection{Implementability}
Refers to the ease with which a security solution can be developed and incorporated into UAV systems. 
\begin{itemize}[noitemsep]
    \item \textbf{High (\fullcirc)} implementability require extensive development time, advanced programming knowledge, and specialized tools. High implementability means the security measure requires fewer resources and is easier to adopt.
    \item \textbf{Medium (\halfcirc)} refers to the moderately implementable mechanisms require reasonable technical expertise and effort to integrate but do not necessitate major architectural changes.
    \item \textbf{Low (\emptycirc)} defines the security mechanisms with low implementability are easy to develop and deploy, requiring minimal technical expertise. 
\end{itemize}
   
\subsubsection{Robustness}
The ability of a security mechanism to withstand attacks and function correctly under adverse conditions. A robust system remains secure despite variations in operational environments and attack attempts.
\begin{itemize}[noitemsep]
    \item \textbf{High~(\fullcirc)}: High robust mechanisms protect against various attack vectors, ensuring overall system integrity even in adverse conditions.
    \item \textbf{Medium~(\halfcirc)}: Medium robust mechanisms offer protection against common threats but may be vulnerable to sophisticated attacks, such as signature-based IDS and hash-based authentication. 

    \item \textbf{low~(\emptycirc)}: Low robustness  provide limited protection and may only defend against basic attacks, including simple password authentication and static access control lists.
\end{itemize}
\begin{sidewaystable*}[htbp]
\scriptsize
\centering
\caption{Qualitative comparison of the countermeasures in UAV Systems in terms of their implementation overhead and performance efficiency.}
\label{Table:Countermeasures}
\begin{tabular}{|l|l|ccc|ccc|cccc|c|c|c|c|}
\hline
\multicolumn{1}{|c|}{}                                                                                                                      & \multicolumn{1}{c|}{}                                                                                                        & \multicolumn{3}{c|}{\textbf{Impact}}                                           & \multicolumn{3}{c|}{\textbf{Model}}                                            & \multicolumn{4}{c|}{\textbf{Implementation Overhead}}                                                                                 &                                    &                                    &                                    &                                      \\ \cline{3-12}
\multicolumn{1}{|c|}{\multirow{-2}{*}{\textbf{Category}}}                                                                                   & \multicolumn{1}{c|}{\multirow{-2}{*}{\textbf{Method}}}                                                                       & \multicolumn{1}{c|}{\textbf{C}} & \multicolumn{1}{c|}{\textbf{I}} & \textbf{A} & \multicolumn{1}{c|}{\textbf{P}} & \multicolumn{1}{c|}{\textbf{D}} & \textbf{R} & \multicolumn{1}{c|}{\textbf{Comput.}} & \multicolumn{1}{c|}{\textbf{Memory}} & \multicolumn{1}{c|}{\textbf{Latency}} & \textbf{Comm.} & \multirow{-2}{*}{\textbf{Deploy.}} & \multirow{-2}{*}{\textbf{Implem.}} & \multirow{-2}{*}{\textbf{Robust.}} & \multirow{-2}{*}{\textbf{Usability}} \\ \hline
                                                                                                                      & {C1:~Hash Function-based Authentication~\cite{hu2021random}}                          & \multicolumn{1}{c|}{{\color[HTML]{FE0000} \xmark}}          & \multicolumn{1}{c|}{\cmark}          & {\color[HTML]{FE0000} \xmark}          & \multicolumn{1}{c|}{\cmark}          & \multicolumn{1}{c|}{{\color[HTML]{FE0000} \xmark}}          & {\color[HTML]{FE0000} \xmark}          & \multicolumn{1}{c|}{\emptycirc}                & \multicolumn{1}{c|}{\emptycirc}               & \multicolumn{1}{c|}{\emptycirc}                & \emptycirc              & \fullcirc                                  & \fullcirc                                  & \halfcirc                                  & \fullcirc                                    \\ \cline{2-16} 
                                                                                                                      & {C2:~  Elliptic Curve Cryptography based Authentication~\cite{el2022authenticating}}                                                      & \multicolumn{1}{c|}{\cmark}          & \multicolumn{1}{c|}{\cmark}          & {\color[HTML]{FE0000} \xmark}          & \multicolumn{1}{c|}{\cmark}          & \multicolumn{1}{c|}{{\color[HTML]{FE0000} \xmark}}          & {\color[HTML]{FE0000} \xmark}          & \multicolumn{1}{c|}{\halfcirc}                & \multicolumn{1}{c|}{\halfcirc}               & \multicolumn{1}{c|}{\halfcirc}                & \emptycirc              & \fullcirc                                  & \halfcirc                                  & \fullcirc                                  & \fullcirc                                    \\ \cline{2-16} 
                                                                                                                      & {C3:~SHA-256 and XOR-based Auth \& Encryption~\cite{ismael2021authentication}}                                                & \multicolumn{1}{c|}{\cmark}          & \multicolumn{1}{c|}{\cmark}          & {\color[HTML]{FE0000} \xmark}          & \multicolumn{1}{c|}{\cmark}          & \multicolumn{1}{c|}{{\color[HTML]{FE0000} \xmark}}          & {\color[HTML]{FE0000} \xmark}          & \multicolumn{1}{c|}{\halfcirc}                & \multicolumn{1}{c|}{\halfcirc}               & \multicolumn{1}{c|}{\halfcirc}                & \halfcirc              & \fullcirc                                  & \halfcirc                                  & \halfcirc                                  & \fullcirc                                    \\ \cline{2-16} 
                                                                                                                      & {C4:  Blockchain based Authentication~\cite{akram2023blockchain}}                                                                       & \multicolumn{1}{c|}{\cmark}          & \multicolumn{1}{c|}{\cmark}          & \cmark          & \multicolumn{1}{c|}{\cmark}          & \multicolumn{1}{c|}{{\color[HTML]{FE0000} \xmark}}          & \cmark          & \multicolumn{1}{c|}{\fullcirc}                & \multicolumn{1}{c|}{\fullcirc}               & \multicolumn{1}{c|}{\fullcirc}                & \fullcirc              & \emptycirc                                  & \emptycirc                                  & \fullcirc                                  & \emptycirc                                    \\ \cline{2-16} 
                                                                                                                      & {C5:~   Biometric based Authentication~\cite{gautam2024probably}}                                                                        & \multicolumn{1}{c|}{\cmark}          & \multicolumn{1}{c|}{\cmark}          & {\color[HTML]{FE0000} \xmark}          & \multicolumn{1}{c|}{\cmark}          & \multicolumn{1}{c|}{{\color[HTML]{FE0000} \xmark}}          & {\color[HTML]{FE0000} \xmark}          & \multicolumn{1}{c|}{\fullcirc}                & \multicolumn{1}{c|}{\fullcirc}               & \multicolumn{1}{c|}{\fullcirc}                & \emptycirc              & \emptycirc                                  & \emptycirc                                  & \fullcirc                                  & \halfcirc                                    \\ \cline{2-16} 
                                                                                                                      & {C6:~   Homomorphic Encryption~\cite{yan2024secure}}                                     & \multicolumn{1}{c|}{\cmark}          & \multicolumn{1}{c|}{\cmark}          & {\color[HTML]{FE0000} \xmark}          & \multicolumn{1}{c|}{\cmark}          & \multicolumn{1}{c|}{{\color[HTML]{FE0000} \xmark}}          & {\color[HTML]{FE0000} \xmark}          & \multicolumn{1}{c|}{\fullcirc}                & \multicolumn{1}{c|}{\fullcirc}               & \multicolumn{1}{c|}{\fullcirc}                & \fullcirc              & \emptycirc                                  & \emptycirc                                  & \fullcirc                                  & \emptycirc                                    \\ \cline{2-16} 
                                                                                                                      & {C7:~   PKI-based Authentication and Encryption~\cite{jadhav2024pki}}                                                               & \multicolumn{1}{c|}{\cmark}          & \multicolumn{1}{c|}{\cmark}          & {\color[HTML]{FE0000} \xmark}          & \multicolumn{1}{c|}{\cmark}          & \multicolumn{1}{c|}{{\color[HTML]{FE0000} \xmark}}          & {\color[HTML]{FE0000} \xmark}          & \multicolumn{1}{c|}{\fullcirc}                & \multicolumn{1}{c|}{\fullcirc}               & \multicolumn{1}{c|}{\fullcirc}                & \halfcirc              & \halfcirc                                  & \halfcirc                                  & \fullcirc                                  & \halfcirc                                    \\ \cline{2-16} 
                                                                                                                      & {C8:~   End-to-End Encryption~\cite{zhang2024auto}}                                      & \multicolumn{1}{c|}{\cmark}          & \multicolumn{1}{c|}{{\color[HTML]{FE0000} \xmark}}          & {\color[HTML]{FE0000} \xmark}          & \multicolumn{1}{c|}{{\color[HTML]{FE0000} \xmark}}          & \multicolumn{1}{c|}{{\color[HTML]{FE0000} \xmark}}          & {\color[HTML]{FE0000} \xmark}          & \multicolumn{1}{c|}{\halfcirc}                & \multicolumn{1}{c|}{\halfcirc}               & \multicolumn{1}{c|}{\halfcirc}                & \halfcirc              & \fullcirc                                  & \fullcirc                                  & \fullcirc                                  & \fullcirc                                    \\ \cline{2-16} 
                                                                                                                      & {C9:~   Encryption based Secure Communication~\cite{shafique2022lightweight}}            & \multicolumn{1}{c|}{\cmark}          & \multicolumn{1}{c|}{{\color[HTML]{FE0000} \xmark}}          & {\color[HTML]{FE0000} \xmark}          & \multicolumn{1}{c|}{{\color[HTML]{FE0000} \xmark}}          & \multicolumn{1}{c|}{{\color[HTML]{FE0000} \xmark}}          & {\color[HTML]{FE0000} \xmark}          & \multicolumn{1}{c|}{\halfcirc}                & \multicolumn{1}{c|}{\halfcirc}               & \multicolumn{1}{c|}{\halfcirc}                & \halfcirc              & \fullcirc                                  & \halfcirc                                  & \halfcirc                                  & \fullcirc                                    \\ \cline{2-16} 
                                                                                                                      & {C10:~Digital Signatures based Secure Communication~\cite{aissaoui2023authenticating}} & \multicolumn{1}{c|}{{\color[HTML]{FE0000} \xmark}}          & \multicolumn{1}{c|}{\cmark}          & {\color[HTML]{FE0000} \xmark}          & \multicolumn{1}{c|}{\cmark}          & \multicolumn{1}{c|}{{\color[HTML]{FE0000} \xmark}}          & {\color[HTML]{FE0000} \xmark}          & \multicolumn{1}{c|}{\halfcirc}                & \multicolumn{1}{c|}{\halfcirc}               & \multicolumn{1}{c|}{\halfcirc}                & \halfcirc              & \halfcirc                                  & \halfcirc                                  & \fullcirc                                  & \halfcirc                                    \\ \cline{2-16} 
                                                                                                                      & {C11:~ECDSA and ECDH based Secure Communication~\cite{ko2021drone}}                                & \multicolumn{1}{c|}{\cmark}          & \multicolumn{1}{c|}{\cmark}          & {\color[HTML]{FE0000} \xmark}          & \multicolumn{1}{c|}{\cmark}          & \multicolumn{1}{c|}{{\color[HTML]{FE0000} \xmark}}          & \cmark          & \multicolumn{1}{c|}{\halfcirc}                & \multicolumn{1}{c|}{\halfcirc}               & \multicolumn{1}{c|}{\halfcirc}                & \halfcirc              & \fullcirc                                  & \fullcirc                                  & \fullcirc                                  & \fullcirc                                    \\ \cline{2-16} 
                                                                                                                      & {C12:~Diffie-Hellman Key Exchange with HMAC~\cite{xia2024quantum}}                                                                 & \multicolumn{1}{c|}{\cmark}          & \multicolumn{1}{c|}{\cmark}          & {\color[HTML]{FE0000} \xmark}          & \multicolumn{1}{c|}{\cmark}          & \multicolumn{1}{c|}{{\color[HTML]{FE0000} \xmark}}          & {\color[HTML]{FE0000} \xmark}          & \multicolumn{1}{c|}{\halfcirc}                & \multicolumn{1}{c|}{\halfcirc}               & \multicolumn{1}{c|}{\halfcirc}                & \halfcirc              & \fullcirc                                  & \fullcirc                                  & \halfcirc                                  & \fullcirc                                    \\ \cline{2-16} 
                                                                                                                      & {C13:~Session Key Establishment~\cite{sc2023secure}}                                                                             & \multicolumn{1}{c|}{\cmark}          & \multicolumn{1}{c|}{\cmark}          & {\color[HTML]{FE0000} \xmark}          & \multicolumn{1}{c|}{\cmark}          & \multicolumn{1}{c|}{{\color[HTML]{FE0000} \xmark}}          & {\color[HTML]{FE0000} \xmark}          & \multicolumn{1}{c|}{\halfcirc}                & \multicolumn{1}{c|}{\halfcirc}               & \multicolumn{1}{c|}{\halfcirc}                & \halfcirc              & \emptycirc                                  & \emptycirc                                  & \fullcirc                                  & \emptycirc                                    \\ \cline{2-16} 
\multirow{-14}{*}{{   \textbf{\begin{tabular}[c]{@{}l@{}}T1: Wireless \\ Communication \\ Module\end{tabular}}}} & {C14:~Secure Multi-Hop Aerial Relay System~\cite{chen2018multiple}}                                                                  & \multicolumn{1}{c|}{\cmark}          & \multicolumn{1}{c|}{{\color[HTML]{FE0000} \xmark}}          & \cmark          & \multicolumn{1}{c|}{\cmark}          & \multicolumn{1}{c|}{{\color[HTML]{FE0000} \xmark}}          & \cmark          & \multicolumn{1}{c|}{\fullcirc}                & \multicolumn{1}{c|}{\fullcirc}               & \multicolumn{1}{c|}{\fullcirc}                & \fullcirc              & \fullcirc                                  & \fullcirc                                  & \halfcirc                                  & \fullcirc                                    \\ \hline
                                                                                                                      & {C15:~Sensor Data Integrity Verification~\cite{jain2017drone}}                                                                    & \multicolumn{1}{c|}{{\color[HTML]{FE0000} \xmark}}          & \multicolumn{1}{c|}{\cmark}          & {\color[HTML]{FE0000} \xmark}          & \multicolumn{1}{c|}{\cmark}          & \multicolumn{1}{c|}{{\color[HTML]{FE0000} \xmark}}          & {\color[HTML]{FE0000} \xmark}          & \multicolumn{1}{c|}{\halfcirc}                & \multicolumn{1}{c|}{\emptycirc}               & \multicolumn{1}{c|}{\emptycirc}                & \emptycirc              & \emptycirc                                  & \emptycirc                                  & \fullcirc                                  & \emptycirc                                    \\ \cline{2-16} 
                                                                                                                      & C16:~Multi-Sensor Fusion based Anomaly Detection~\cite{alzahrani2024enhancing}                                                                                  & \multicolumn{1}{c|}{{\color[HTML]{FE0000} \xmark}}          & \multicolumn{1}{c|}{\cmark}          & \cmark          & \multicolumn{1}{c|}{\cmark}          & \multicolumn{1}{c|}{\cmark}          & \cmark          & \multicolumn{1}{c|}{\fullcirc}                & \multicolumn{1}{c|}{\fullcirc}               & \multicolumn{1}{c|}{\fullcirc}                & \halfcirc              & \halfcirc                                  & \halfcirc                                  & \fullcirc                                  & \halfcirc                                    \\ \cline{2-16} 
                                                                                                                      &C17:~ GPS Signal Authentication~\cite{nair2025towards}                                                                                                    & \multicolumn{1}{c|}{{\color[HTML]{FE0000} \xmark}}          & \multicolumn{1}{c|}{\cmark}          & \cmark          & \multicolumn{1}{c|}{\cmark}          & \multicolumn{1}{c|}{{\color[HTML]{FE0000} \xmark}}          & \cmark          & \multicolumn{1}{c|}{\halfcirc}                & \multicolumn{1}{c|}{\fullcirc}               & \multicolumn{1}{c|}{\fullcirc}                & \halfcirc              & \emptycirc                                  & \emptycirc                                  & \fullcirc                                  & \emptycirc                                    \\ \cline{2-16} 
                                                                                                                      &C18:~ Anomaly Detection for Camera Spoofing~\cite{chriki2020uav}                                             & \multicolumn{1}{c|}{{\color[HTML]{FE0000} \xmark}}          & \multicolumn{1}{c|}{\cmark}          & \cmark          & \multicolumn{1}{c|}{\cmark}          & \multicolumn{1}{c|}{\cmark}          & \cmark          & \multicolumn{1}{c|}{\fullcirc}                & \multicolumn{1}{c|}{\fullcirc}               & \multicolumn{1}{c|}{\fullcirc}                & \halfcirc              & \emptycirc                                  & \emptycirc                                  & \fullcirc                                  & \emptycirc                                    \\ \cline{2-16} 
                                                                                                                      &C18:~ AI-based Outlier Detection in Point Clouds~\cite{guefrachi2024advanced}                                                                                   & \multicolumn{1}{c|}{{\color[HTML]{FE0000} \xmark}}          & \multicolumn{1}{c|}{\cmark}          & \cmark          & \multicolumn{1}{c|}{\cmark}          & \multicolumn{1}{c|}{\cmark}          & \cmark          & \multicolumn{1}{c|}{\fullcirc}                & \multicolumn{1}{c|}{\fullcirc}               & \multicolumn{1}{c|}{\fullcirc}                & \halfcirc              & \emptycirc                                  & \emptycirc                                  & \fullcirc                                  & \emptycirc                                    \\ \cline{2-16} 
                                                                                                                      &C19:~ Real-time Spectrum Monitoring~\cite{han2024real}                                                                                                & \multicolumn{1}{c|}{{\color[HTML]{FE0000} \xmark}}          & \multicolumn{1}{c|}{\cmark}          & \cmark          & \multicolumn{1}{c|}{\cmark}          & \multicolumn{1}{c|}{{\color[HTML]{FE0000} \xmark}}          & \cmark          & \multicolumn{1}{c|}{\fullcirc}                & \multicolumn{1}{c|}{\emptycirc}               & \multicolumn{1}{c|}{\halfcirc}                & \halfcirc              & \fullcirc                                  & \fullcirc                                  & \halfcirc                                  & \fullcirc                                    \\ \cline{2-16} 
                                                                                                                      &C20:~ Frequency Hopping Spread Spectrum (FHSS)~\cite{de2024enhancing}                                                                                     & \multicolumn{1}{c|}{\cmark}          & \multicolumn{1}{c|}{{\color[HTML]{FE0000} \xmark}}          & \cmark          & \multicolumn{1}{c|}{{\color[HTML]{FE0000} \xmark}}          & \multicolumn{1}{c|}{{\color[HTML]{FE0000} \xmark}}          & \cmark          & \multicolumn{1}{c|}{\halfcirc}                & \multicolumn{1}{c|}{\emptycirc}               & \multicolumn{1}{c|}{\halfcirc}                & \halfcirc              & \fullcirc                                  & \fullcirc                                  & \halfcirc                                  & \fullcirc                                    \\ \cline{2-16} 
\multirow{-8}{*}{{   \textbf{T2: Sensors}}}                                                                      &C21:~Direct Sequence Spread Spectrum (DSSS)~\cite{zellal2025spread}                                                                                       & \multicolumn{1}{c|}{\cmark}          & \multicolumn{1}{c|}{{\color[HTML]{FE0000} \xmark}}          & \cmark          & \multicolumn{1}{c|}{{\color[HTML]{FE0000} \xmark}}          & \multicolumn{1}{c|}{{\color[HTML]{FE0000} \xmark}}          & \cmark          & \multicolumn{1}{c|}{\halfcirc}                & \multicolumn{1}{c|}{\emptycirc}               & \multicolumn{1}{c|}{\emptycirc}                & \emptycirc              & \emptycirc                                  & \emptycirc                                  & \fullcirc                                  & \emptycirc                                    \\ \hline
                                                                                                                                            &C22:~Secure Boot/ Hardware Root of Trust (HRoT)~\cite{markantonakis2016secure}                                                                                   & \multicolumn{1}{c|}{{\color[HTML]{FE0000} \xmark}}          & \multicolumn{1}{c|}{\cmark}          & {\color[HTML]{FE0000} \xmark}          & \multicolumn{1}{c|}{{\color[HTML]{FE0000} \xmark}}          & \multicolumn{1}{c|}{\cmark}          & {\color[HTML]{FE0000} \xmark}          & \multicolumn{1}{c|}{\emptycirc}                & \multicolumn{1}{c|}{\halfcirc}               & \multicolumn{1}{c|}{\halfcirc}                & \emptycirc              & \emptycirc                                  & \emptycirc                                  & \fullcirc                                  & \emptycirc                                    \\ \cline{2-16} 
                                                                                                                                            &C23:~Trusted Execution Environments (TEE)~\cite{liao2023unmanned}                                                                                         & \multicolumn{1}{c|}{\cmark}          & \multicolumn{1}{c|}{\cmark}          & {\color[HTML]{FE0000} \xmark}          & \multicolumn{1}{c|}{\cmark}          & \multicolumn{1}{c|}{{\color[HTML]{FE0000} \xmark}}          & \cmark          & \multicolumn{1}{c|}{\halfcirc}                & \multicolumn{1}{c|}{\fullcirc}               & \multicolumn{1}{c|}{\halfcirc}                & \emptycirc              & \emptycirc                                  & \emptycirc                                  & \fullcirc                                  & \emptycirc                                    \\ \cline{2-16} 
                                                                                                                                            &C24:~Hardware Security Modules (HSM)~\cite{pirker2020global}                                                                                             & \multicolumn{1}{c|}{\cmark}          & \multicolumn{1}{c|}{\cmark}          & {\color[HTML]{FE0000} \xmark}          & \multicolumn{1}{c|}{\cmark}          & \multicolumn{1}{c|}{\cmark}          & \cmark          & \multicolumn{1}{c|}{\fullcirc}                & \multicolumn{1}{c|}{\emptycirc}               & \multicolumn{1}{c|}{\emptycirc}                & \emptycirc              & \emptycirc                                  & \emptycirc                                  & \fullcirc                                  & \emptycirc                                    \\ \cline{2-16} 
                                                                                                                                            &C25:~Physical Unclonable Functions (PUFs)~\cite{alkatheiri2022lightweight}                                                                                         & \multicolumn{1}{c|}{\cmark}          & \multicolumn{1}{c|}{\cmark}          & {\color[HTML]{FE0000} \xmark}          & \multicolumn{1}{c|}{\cmark}          & \multicolumn{1}{c|}{\cmark}          & {\color[HTML]{FE0000} \xmark}          & \multicolumn{1}{c|}{\emptycirc}                & \multicolumn{1}{c|}{\halfcirc}               & \multicolumn{1}{c|}{\halfcirc}                & \emptycirc              & \emptycirc                                  & \emptycirc                                  & \fullcirc                                  & \emptycirc                                    \\ \cline{2-16} 
\multirow{-5}{*}{{\textbf{\begin{tabular}[c]{@{}l@{}}T3: Hardware \\ and\\ Firmware\end{tabular}}}}
                                                                                           &C26:~PQC based Hardware Accelerators~\cite{amiet2018fpga}                                                                        & \multicolumn{1}{c|}{\cmark}          & \multicolumn{1}{c|}{\cmark}          & {\color[HTML]{FE0000} \xmark}          & \multicolumn{1}{c|}{\cmark}          & \multicolumn{1}{c|}{\cmark}          & {\color[HTML]{FE0000} \xmark}          & \multicolumn{1}{c|}{\fullcirc}                & \multicolumn{1}{c|}{\halfcirc}               & \multicolumn{1}{c|}{\halfcirc}                & \emptycirc              & \fullcirc                                  & \fullcirc                                  & \halfcirc                                  & \fullcirc                                    \\ \hline
                                                                                                                                            &C27:~ Role-Based Access Control (RBAC)~\cite{jeong2012rbac}                                                                                            & \multicolumn{1}{c|}{\cmark}          & \multicolumn{1}{c|}{{\color[HTML]{FE0000} \xmark}}          & {\color[HTML]{FE0000} \xmark}          & \multicolumn{1}{c|}{\cmark}          & \multicolumn{1}{c|}{{\color[HTML]{FE0000} \xmark}}          & \cmark          & \multicolumn{1}{c|}{\emptycirc}                & \multicolumn{1}{c|}{\emptycirc}               & \multicolumn{1}{c|}{\emptycirc}                & \emptycirc              & \halfcirc                                  & \halfcirc                                  & \halfcirc                                  & \halfcirc                                    \\ \cline{2-16} 
                                                                                                                                            &C28:~Mandatory Access Control~\cite{beck2020privaros}                                                       & \multicolumn{1}{c|}{\cmark}          & \multicolumn{1}{c|}{{\color[HTML]{FE0000} \xmark}}          & {\color[HTML]{FE0000} \xmark}          & \multicolumn{1}{c|}{\cmark}          & \multicolumn{1}{c|}{{\color[HTML]{FE0000} \xmark}}          & \cmark          & \multicolumn{1}{c|}{\halfcirc}                & \multicolumn{1}{c|}{\halfcirc}               & \multicolumn{1}{c|}{\halfcirc}                & \halfcirc              & \halfcirc                                  & \halfcirc                                  & \halfcirc                                  & \halfcirc                                    \\ \cline{2-16} 
                                                                                                                                            &C29:~Attribute-Based Access Control (ABAC)~\cite{japp2024fly}                                                                                        & \multicolumn{1}{c|}{\cmark}          & \multicolumn{1}{c|}{{\color[HTML]{FE0000} \xmark}}          & {\color[HTML]{FE0000} \xmark}          & \multicolumn{1}{c|}{\cmark}          & \multicolumn{1}{c|}{{\color[HTML]{FE0000} \xmark}}          & \cmark          & \multicolumn{1}{c|}{\halfcirc}                & \multicolumn{1}{c|}{\fullcirc}               & \multicolumn{1}{c|}{\fullcirc}                & \halfcirc              & \fullcirc                                  & \fullcirc                                  & \emptycirc                                  & \fullcirc                                    \\ \cline{2-16} 
                                                                                                                                            &C30:~ Access Control Lists (ACLs)                                                                                                  & \multicolumn{1}{c|}{\cmark}          & \multicolumn{1}{c|}{{\color[HTML]{FE0000} \xmark}}          & {\color[HTML]{FE0000} \xmark}          & \multicolumn{1}{c|}{\cmark}          & \multicolumn{1}{c|}{{\color[HTML]{FE0000} \xmark}}          & \cmark          & \multicolumn{1}{c|}{\emptycirc}                & \multicolumn{1}{c|}{\halfcirc}               & \multicolumn{1}{c|}{\halfcirc}                & \halfcirc              & \emptycirc                                  & \emptycirc                                  & \fullcirc                                  & \emptycirc                                    \\ \cline{2-16} 
                                                                                                                                            &C31:~ Microkernel~\cite{koo2020secure}                                                                                                                  & \multicolumn{1}{c|}{{\color[HTML]{FE0000} \xmark}}          & \multicolumn{1}{c|}{\cmark}          & \cmark          & \multicolumn{1}{c|}{\cmark}          & \multicolumn{1}{c|}{{\color[HTML]{FE0000} \xmark}}          & \cmark          & \multicolumn{1}{c|}{\halfcirc}                & \multicolumn{1}{c|}{\halfcirc}               & \multicolumn{1}{c|}{\emptycirc}                & \emptycirc              & \emptycirc                                  & \emptycirc                                  & \fullcirc                                  & \emptycirc                                    \\ \cline{2-16} 
                                                                                                                                            &C32:~Hypervisors/ Virtual Machine Monitor~\cite{cittadini2023supporting}                                                                                         & \multicolumn{1}{c|}{{\color[HTML]{FE0000} \xmark}}          & \multicolumn{1}{c|}{\cmark}          & \cmark          & \multicolumn{1}{c|}{\cmark}          & \multicolumn{1}{c|}{{\color[HTML]{FE0000} \xmark}}          & \cmark          & \multicolumn{1}{c|}{\fullcirc}                & \multicolumn{1}{c|}{\fullcirc}               & \multicolumn{1}{c|}{\fullcirc}                & \halfcirc              & \fullcirc                                  & \fullcirc                                  & \halfcirc                                  & \fullcirc                                    \\ \cline{2-16} 
                                                                                                                                            &C33:~Secure Patch Updates~\cite{seo2023blockchain}                                                                                                         & \multicolumn{1}{c|}{{\color[HTML]{FE0000} \xmark}}          & \multicolumn{1}{c|}{\cmark}          & \cmark          & \multicolumn{1}{c|}{\cmark}          & \multicolumn{1}{c|}{{\color[HTML]{FE0000} \xmark}}          & \cmark          & \multicolumn{1}{c|}{\halfcirc}                & \multicolumn{1}{c|}{\fullcirc}               & \multicolumn{1}{c|}{\fullcirc}                & \fullcirc              & \halfcirc                                  & \halfcirc                                  & \fullcirc                                  & \halfcirc                                    \\ \cline{2-16} 
                                                                                                                                            &C34:~Secure Bootloader                                                                                                            & \multicolumn{1}{c|}{{\color[HTML]{FE0000} \xmark}}          & \multicolumn{1}{c|}{\cmark}          & {\color[HTML]{FE0000} \xmark}          & \multicolumn{1}{c|}{\cmark}          & \multicolumn{1}{c|}{{\color[HTML]{FE0000} \xmark}}          & \cmark          & \multicolumn{1}{c|}{\halfcirc}                & \multicolumn{1}{c|}{\fullcirc}               & \multicolumn{1}{c|}{\fullcirc}                & \fullcirc              & \halfcirc                                  & \halfcirc                                  & \fullcirc                                  & \halfcirc                                    \\ \cline{2-16} 
\multirow{-9}{*}{\textbf{\begin{tabular}[c]{@{}l@{}}T4: Operating \\Systems and\\ Applications\end{tabular}}}                                         &C35:~Malware Detection~\cite{niu2020malware}                                                                & \multicolumn{1}{c|}{{\color[HTML]{FE0000} \xmark}}          & \multicolumn{1}{c|}{\cmark}          & \cmark          & \multicolumn{1}{c|}{\cmark}          & \multicolumn{1}{c|}{\cmark}          & {\color[HTML]{FE0000} \xmark}          & \multicolumn{1}{c|}{\fullcirc}                & \multicolumn{1}{c|}{\halfcirc}               & \multicolumn{1}{c|}{\halfcirc}                & \halfcirc              & \emptycirc                                  & \emptycirc                                  & \fullcirc                                  & \emptycirc                                    \\ \hline
                                                                                                                                            &C36:~SDN-based Methods~\cite{karegar2025uav}                                                                & \multicolumn{1}{c|}{\cmark}          & \multicolumn{1}{c|}{\cmark}          & \cmark          & \multicolumn{1}{c|}{\cmark}          & \multicolumn{1}{c|}{{\color[HTML]{FE0000} \xmark}}          & {\color[HTML]{FE0000} \xmark}          & \multicolumn{1}{c|}{\fullcirc}                & \multicolumn{1}{c|}{\halfcirc}               & \multicolumn{1}{c|}{\halfcirc}                & \halfcirc              & \emptycirc                                  & \emptycirc                                  & \fullcirc                                  & \emptycirc                                    \\ \cline{2-16} 
                                                                                                                                            &C37:~Blockchain~\cite{hafeez2023blockchain}                                                                 & \multicolumn{1}{c|}{\cmark}          & \multicolumn{1}{c|}{\cmark}          & \cmark          & \multicolumn{1}{c|}{\cmark}          & \multicolumn{1}{c|}{\cmark}          & {\color[HTML]{FE0000} \xmark}          & \multicolumn{1}{c|}{\fullcirc}                & \multicolumn{1}{c|}{\halfcirc}               & \multicolumn{1}{c|}{\halfcirc}                & \halfcirc              & \halfcirc                                  & \halfcirc                                  & \fullcirc                                  & \halfcirc                                    \\ \cline{2-16} 
                                                                                                                                            &C38:~ Network Access Control~\cite{araghizadeh2016efficient}                                                 & \multicolumn{1}{c|}{\cmark}          & \multicolumn{1}{c|}{\cmark}          & {\color[HTML]{FE0000} \xmark}          & \multicolumn{1}{c|}{\cmark}          & \multicolumn{1}{c|}{{\color[HTML]{FE0000} \xmark}}          & {\color[HTML]{FE0000} \xmark}          & \multicolumn{1}{c|}{\halfcirc}                & \multicolumn{1}{c|}{\halfcirc}               & \multicolumn{1}{c|}{\halfcirc}                & \halfcirc              & \halfcirc                                  & \halfcirc                                  & \fullcirc                                  & \halfcirc                                    \\ \cline{2-16} 
                                                                                                                                            &C39:~HECC based Authentication ~\cite{gnanaraj2024hyperelliptic}                                               & \multicolumn{1}{c|}{\cmark}          & \multicolumn{1}{c|}{\cmark}          & {\color[HTML]{FE0000} \xmark}          & \multicolumn{1}{c|}{\cmark}          & \multicolumn{1}{c|}{{\color[HTML]{FE0000} \xmark}}          & {\color[HTML]{FE0000} \xmark}          & \multicolumn{1}{c|}{\halfcirc}                & \multicolumn{1}{c|}{\halfcirc}               & \multicolumn{1}{c|}{\halfcirc}                & \halfcirc              & \halfcirc                                  & \halfcirc                                  & \halfcirc                                  & \halfcirc                                    \\ \cline{2-16} 
                                                                                                                                            &C40:~Intrusion Detection Systems~\cite{senturk2025artificial}                                               & \multicolumn{1}{c|}{{\color[HTML]{FE0000} \xmark}}          & \multicolumn{1}{c|}{\cmark}          & \cmark          & \multicolumn{1}{c|}{{\color[HTML]{FE0000} \xmark}}          & \multicolumn{1}{c|}{\cmark}          & {\color[HTML]{FE0000} \xmark}          & \multicolumn{1}{c|}{\halfcirc}                & \multicolumn{1}{c|}{\halfcirc}               & \multicolumn{1}{c|}{\halfcirc}                & \halfcirc              & \halfcirc                                  & \halfcirc                                  & \halfcirc                                  & \halfcirc                                    \\ \cline{2-16} 
                                                                                                                                            &C41:~Intrusion Prevention Systems~\cite{ntizikira2023secure}                                                & \multicolumn{1}{c|}{{\color[HTML]{FE0000} \xmark}}          & \multicolumn{1}{c|}{\cmark}          & \cmark          & \multicolumn{1}{c|}{\cmark}          & \multicolumn{1}{c|}{{\color[HTML]{FE0000} \xmark}}          & {\color[HTML]{FE0000} \xmark}          & \multicolumn{1}{c|}{\halfcirc}                & \multicolumn{1}{c|}{\halfcirc}               & \multicolumn{1}{c|}{\halfcirc}                & \halfcirc              & \halfcirc                                  & \halfcirc                                  & \fullcirc                                  & \halfcirc                                    \\ \cline{2-16} 
                                                                                                                                            &C42:~Firewalls~\cite{shangte2021research}                                                                   & \multicolumn{1}{c|}{\cmark}          & \multicolumn{1}{c|}{\cmark}          & \cmark          & \multicolumn{1}{c|}{\cmark}          & \multicolumn{1}{c|}{{\color[HTML]{FE0000} \xmark}}          & {\color[HTML]{FE0000} \xmark}          & \multicolumn{1}{c|}{\halfcirc}                & \multicolumn{1}{c|}{\halfcirc}               & \multicolumn{1}{c|}{\emptycirc}                & \halfcirc              & \fullcirc                                  & \fullcirc                                  & \halfcirc                                  & \fullcirc                                    \\ \cline{2-16} 
                                                                                                                                            &C43:~Zero-Knowledge Proofs~\cite{koulianos2024enhancing}                                                    & \multicolumn{1}{c|}{\cmark}          & \multicolumn{1}{c|}{{\color[HTML]{FE0000} \xmark}}          & {\color[HTML]{FE0000} \xmark}          & \multicolumn{1}{c|}{\cmark}          & \multicolumn{1}{c|}{{\color[HTML]{FE0000} \xmark}}          & {\color[HTML]{FE0000} \xmark}          & \multicolumn{1}{c|}{\fullcirc}                & \multicolumn{1}{c|}{\fullcirc}               & \multicolumn{1}{c|}{\fullcirc}                & \halfcirc              & \emptycirc                                  & \emptycirc                                  & \fullcirc                                  & \emptycirc                                    \\ \cline{2-16} 
\multirow{-9}{*}{\textbf{T5: Networks}}                                                                                             &C44:~GCNN based Fraud Detection~\cite{lakhan2022its}                                                                                                          & \multicolumn{1}{c|}{{\color[HTML]{FE0000} \xmark}}          & \multicolumn{1}{c|}{\cmark}          & \cmark          & \multicolumn{1}{c|}{{\color[HTML]{FE0000} \xmark}}          & \multicolumn{1}{c|}{\cmark}          & {\color[HTML]{FE0000} \xmark}          & \multicolumn{1}{c|}{\fullcirc}                & \multicolumn{1}{c|}{\fullcirc}               & \multicolumn{1}{c|}{\fullcirc}                & \halfcirc              & \halfcirc                                  & \halfcirc                                  & \fullcirc                                  & \halfcirc                                    \\ \hline
\end{tabular}
\begin{tablenotes}
    \item \text{Note: Where, C = Confidentiality, I = Integrity, A = Availability, P = Prevention, D = Detection, R = Response, Comput. Overhead = Computation Overhead, Comm. Overhead = Communication Overhead,}
    \item \text{Deploy. = Deployability, Implem. = Implementability, Robust. = Robustness, \emptycirc~= Low, \halfcirc~= Medium, \fullcirc~= High, PQC~= Post-Quantum Cryptography, HECC~= Hyperelliptic Curve Cryptography,}
    \item \text{GCNN~=Graph Convolution Neural Network}
\end{tablenotes}
\end{sidewaystable*}

\subsubsection{Usability}
The ease with which operators and systems can implement and interact with security mechanisms. High usability ensures that security does not hinder UAV operations and is intuitive for users. 
\begin{itemize}[noitemsep]
    \item \textbf{High (\fullcirc)}: High usability denotes security mechanisms that are automated, user-friendly, and require minimal operator intervention.
    \item \textbf{Medium (\halfcirc)}: Medium usability signifies moderately usable mechanisms that require some user interaction for configuration and operation but remain manageable.
    \item \textbf{Low (\emptycirc)}: Low usability indicates complex security mechanisms, requiring extensive manual setup and technical knowledge.
\end{itemize}

\subsection{Discussion on Promising Countermeasures}
After carefully categorizing the countermeasures, we have identified several critical measures that can significantly enhance the security of drone systems as a whole. These countermeasures stand out as vital safeguards, ensuring robust protection and reliability across all UAV systems.
\\
\noindent
\textbf{Encryption of the Communication Link: }
UAV network communication links are vulnerable to attacks, but encrypting messages can ensure confidentiality and prevent data tampering, MITM, and eavesdropping. Symmetric key encryption is preferred due to its speed, and a hybrid method combining RSA or ECC for key exchange with AES for message encryption offers optimal security. Ismael et al. developed a Chaotic map and lightweight HIGHT algorithm for authentication and encryption in UAV networks. Additionally, Atov et al. created secure encryption using OTP, while Podhradsky et al. focused on encrypting radio control links. 
\\
\noindent
\textbf{Digital Signatures and Certificates: }
Digital signatures authenticate UAV messages, ensuring data integrity and non-repudiation. Certificates within a PKI validate UAV identities, enabling encrypted communication and trust establishment. They prevent data tampering and unauthorized access, ensuring regulatory compliance. Short-certificate proxy signature schemes improve efficiency. Verma et al. \cite{verma2019cb} suggested a short-certificate-based proxy signature scheme.
\\
\noindent
\textbf{Secure Communication Protocols: }
UAV networks employ TLS/SSL for encrypted transmission, IPsec for secure packet exchange, SSH for remote access, and VPNs for safe communication over untrusted networks. Digital signatures are also employed in designing secure communication protocol which further enhance trust. Khan et al. \cite{khan2020emerging} explored emerging secure communication schemes in UAV communication protocols. 
\\
\noindent
\textbf{User Authentication and Authorization: }
Authentication ensures only verified users (e.g., drone operators) access UAV systems, using credentials, biometrics, or multi-factor authentication. Authorization defines user privileges, restricting unauthorized access. Alladi et al. \cite{alladi2021drone} proposed a secure authentication scheme for drones.
\\
\noindent
\textbf{Device Authentication: }  
Device authentication in UAV networks verifies the identity and authenticity of drones using cryptographic methods like digital signatures. Each drone has a unique identifier, and an authentication server maintains a database of authorized drones and their associated keys to ensure only verified drones can operate and access sensitive data. Successfully authenticated drones establish secure communication channels to maintain data integrity and confidentiality. Gope et al. \cite{gope2021provably} proposed a device authentication mechanism for UAVs. In swarm scenarios, authentication becomes more complex due to the network's dynamic nature. Group authentication verifies the swarm's integrity, while dynamic authentication updates the list of verified drones.
\\
\noindent
\textbf{Blockchain Technology:}  
Blockchain secures UAV operations by preventing data tampering, encrypting communication, and automating access control via smart contracts. Rana et al. \cite{rana2019intelligent} designed a blockchain-based scheme using SHA-3 hashing, time-stamping, and encryption to verify data integrity in UAV systems. 
\section{Potential Research Directions}\label{sec7_Research_Directions}
In the above two sections, we highlight the importance of studying drone systems, particularly in understanding attacks and countermeasures, categorizing them, and mapping their impacts. This systematization of attacks and countermeasures reveals the following research gaps in research that can be undertaken in future. 

\subsection{FCS Robustness}
The future research should aim to facilitate strict system level policy enforcements in the drone's hardware and software level to tackle various system level attacks as discussed in the Section~\ref{sec4_Systematization_of_Attacks}. It is imperative to investigate unexplored vulnerabilities in the drone system, which must be identified to protect the system from potential zero-day attacks. 
To highlights the gaps in the hardware and software level attacks and countermeasures, we have outlined some research challenges as follows:

\subsubsection{Potential Attacks}
\noindent
\textbf{Unsigned Firmware Updates:} If firmware updates are not signed, attackers can run malicious code on the UAV system. This allows them to alter flight behavior, disable safety features, or retrieve sensitive data. Such attacks can be carried out remotely if the update is not authenticated.
\\
\noindent 
\textbf{Insecure Configuration Storage:} Storing flight parameters, security keys and other critical information in unprotected memory can allow attackers to modify or delete them. Changes in these values can cause unsafe flight behavior or disable safety critical parameters. Hence, we must store the critical information of the drone, in the secure storage and must perform integrity checks, before the system accept and run with  malicious configurations. 
\\
\noindent 
\textbf{Adaptive Sensor Spoofing:} Attackers can feed false data to sensors such as GPS, IMU, or magnetometers in a way that adapts to the UAV’s movement. This can mislead the navigation system into deviating from its planned path. Adaptive spoofing can bypass simple sensor validation mechanisms and cause controlled crashes or hijacks.
\\
\noindent
\textbf{Hardware Supply Chain Tampering:} Malicious components or modified firmware can be introduced into the UAV during manufacturing or distribution. These backdoors may remain dormant until activated during flight. Detecting such tampering is difficult without strict supply chain security measures.
\\
\noindent
\textbf{Cross-layer Fault Injection:} By exploiting interactions between hardware, firmware, and software, attackers can trigger faults that cascade through multiple layers of the system. Techniques like voltage glitching or electromagnetic interference can cause unintended behaviors. Such attacks can bypass traditional single-layer defenses
\subsubsection{Potential Countermeasures}
\noindent
\textbf{Firmware Integrity and Attestation:} We should regularly verify UAV firmware to detect any unauthorized modifications or malware. This can be done locally on the drone or remotely from a ground station. A Trusted Execution Environment (TEE), such as ARM TrustZone, can securely store keys and run integrity checks in an isolated secure world. 
\\
\noindent
\textbf{Secure Boot and Root of Trust:} We should ensure that the UAV only runs authenticated firmware from startup. A hardware-based root of trust can verify the bootloader and operating system before they are loaded. This prevents attackers from running malicious code at the earliest stage of operation.
\\
\noindent
\textbf{TEE-Based Secure Storage:} We should use a TEE to protect sensitive data such as encryption keys, authentication credentials, and critical flight parameters. The secure storage is isolated from the main operating system, making it resistant to tampering. This ensures that even if the main firmware is compromised, critical secrets remain safe.
\\
\noindent
\textbf{Redundant Sensor Data:} We should combine data from multiple sensors to maintain reliable navigation even if one sensor is compromised. For example, integrate encrypted GPS with inertial measurement units (IMUs) and visual odometry. This multi-sensor fusion can reduce the impact of GPS spoofing or jamming attacks.

\subsection{GCS Robustness} 
\subsubsection{Potential Attacks}
\noindent
\textbf{Interface-level Exploits:} Debug ports such as UART, JTAG, or SWD left accessible on the UAV can allow direct access to firmware and memory. Attackers can use these interfaces to extract sensitive information or overwrite system code. Physical access to the UAV greatly increases the risk of such exploits.
\\
\noindent   
\textbf{Side-channel Attacks on GCS Hardware:} By analyzing physical signals such as power consumption, electromagnetic emissions, or timing patterns from the ground control station, attackers can deduce sensitive information like encryption keys. These attacks can bypass software-level protections entirely. Shielding and hardware hardening are required to mitigate such risks.

\subsubsection{Potential Countermeasures} 
\noindent
\textbf{Operator Command Verification via TEE:} All commands sent from the ground control station should be cryptographically signed and verified before execution. A Trusted Execution Environment (TEE) such as ARM TrustZone can securely store signing keys and perform verification in isolation from the main flight software. This prevents unauthorized or tampered commands from influencing UAV behavior.
\\
\noindent  
\textbf{Isolated Control Sandboxing:} Critical drone command modules can be executed within containerized or virtualized environments on the UAV. This limits the impact of any compromise by isolating malicious code from the main control system. Sandboxing ensures that even if a module is exploited, it cannot directly access or damage core flight functions.

\subsection{FCS-GCS Communication Robustness}
Next, we focus on the discussion to ensure that drone and GCS communication, as well as drone-to-drone communication, must be secure.
The communication system should address the following key challenges:

\subsubsection{Potential Attacks}
\noindent
\textbf{Unencrypted telemetry/control links:} If communication between the UAV and the ground control station is unencrypted, attackers can intercept and read sensitive data. They can also inject malicious commands to alter flight behavior or disable the UAV. Such vulnerabilities are especially dangerous when drones operate over long ranges.
\\
\noindent    
\textbf{Protocol Fuzzing:} By sending malformed or unexpected data packets to the UAV, attackers can identify and exploit weaknesses in its communication protocols. This can lead to system crashes, denial of service, or even remote code execution. Automated fuzzing tools make it easier for attackers to discover these flaws quickly.
\\
\noindent   
\textbf{Man-in-the-middle (MITM) attacks:} In a MITM scenario, attackers position themselves between the UAV and the ground control station to intercept and alter messages. This enables them to manipulate telemetry data, inject false commands, or block legitimate instructions. Without strong authentication and encryption, detecting such attacks can be challenging.
\\
\noindent
\textbf{Multi-drone Command Injection:} In swarm operations, attackers can compromise the communication channel between drones to inject false commands. This can cause drones to collide, disperse, or abandon their mission. Without robust authentication and integrity checks, a single compromised node can disrupt the entire swarm.
\\
\noindent
\textbf{Protocol Downgrade Attacks:} Attackers can force the UAV or ground control station to switch to an older, less secure communication protocol. This can bypass modern encryption or authentication mechanisms, exposing the link to eavesdropping and command injection. Such attacks are particularly dangerous when legacy protocol support is left enabled for compatibility.
\\
\noindent    
\textbf{Cross-link Replay Attacks:} In multi-drone or relay-based communication setups, attackers can capture valid commands from one link and replay them on another. This can trigger unintended actions such as route changes, payload release, or emergency landings. Without freshness checks or sequence numbers, the UAV cannot distinguish between a legitimate new command and a replayed one.
\\
\noindent   
\textbf{Swarm Coordination Exploits:} In coordinated multi-drone operations, attackers can target the swarm control algorithm to cause de-synchronization or collisions. By injecting false status messages or altering position reports, they can disrupt formation flying and mission execution. Compromise of a single node can lead to cascading failures across the entire swarm.

\subsubsection{Potential Countermeasures} 
\noindent
\textbf{Identity Authentication of Drone and GCS:} Each UAV and ground control station should have a unique cryptographic identity to prevent impersonation. Mutual authentication protocols can ensure that both ends verify each other before any control or data exchange. This prevents rogue devices from joining or hijacking UAV operations.
\\
\noindent  
\textbf{Continuous Network Path Integrity Checking:} The UAV and ground control station should continuously monitor the communication path for signs of interception or rerouting. Cryptographic integrity verification and route validation can help detect man-in-the-middle or path manipulation attacks. This ensures that data exchanged between UAVs and controllers remains authentic and unaltered in real time.
\\
\noindent
\textbf{Lightweight Secure Communication Protocol:} Lack of authenticated encryption in UAV communication channels exposes them to man-in-the-middle (MITM) and spoofing attacks. Integrating end-to-end authenticated encryption mechanisms ensures confidentiality, integrity, and authenticity of messages. Using lightweight cryptographic algorithms minimizes performance overhead, making them suitable for resource-constrained drone hardware.
\\
\noindent   
\textbf{Strong Authentication Mechanisms:} Secure authentication of UAVs and their operators can be enforced through multi-factor authentication (MFA) combined with robust cryptographic techniques. This prevents unauthorized access even if one authentication factor is compromised. Such mechanisms add an extra layer of defense against credential theft and brute-force attacks.

\section{Establishment of Drone-Regulator}

\label{sec6_Security_Policies_for_Regulating_Drones}

\begin{figure*}[ht]
    \centering
    \includegraphics[width=0.5\linewidth]{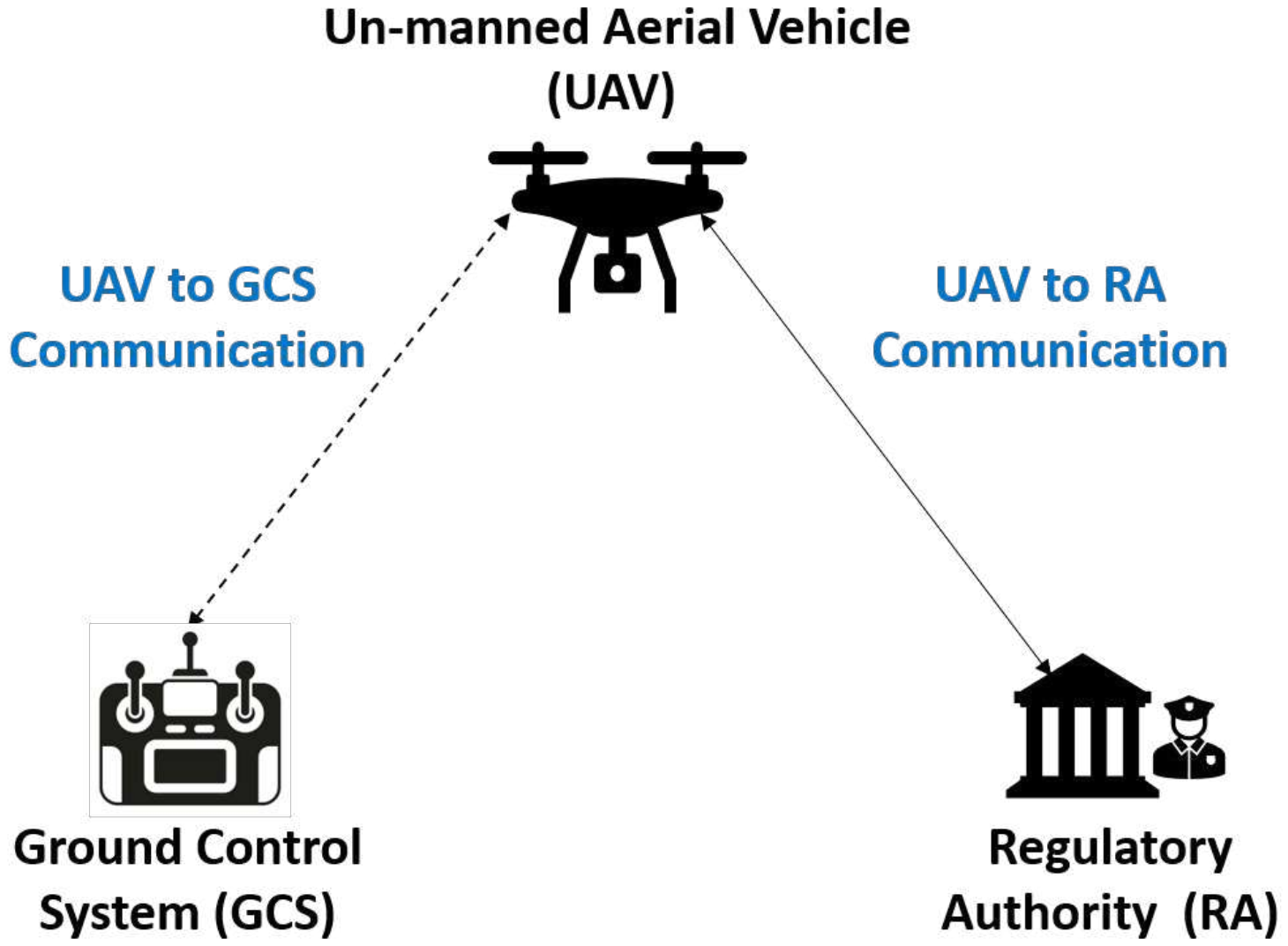}
    \caption{Considered drone system consisting of three entities: the drone, the Ground Control System (GCS), and a proposed \textit{Regulatory Authority} introduced for policy-enforcement.}   
    \label{fig:considered_system}
\end{figure*}

From the above analysis of attacks against drones and their countermeasures, we note that there are potential unmitigated vulnerabilities in the UAV networks that can be exploited to compromise these flying vehicles. This motivates us to explore countermeasures that prevent drones from launching attacks. Specifically, we explore the security-based policy enforcement for drone networks. 
\par
International aviation authorities play a crucial role in regulating drone operations. The Federal Aviation Administration (FAA)~\cite{hatfield2020integration} in the United States enforces regulations that mandate drone registration, airspace classification, and Remote ID tracking. It also oversees pilot certification and beyond-visual line-of-sight (BVLOS) operations. Similarly, the European Union Aviation Safety Agency (EASA)~\cite{europaEasyAccess, rushiti2024overview} and the Civil Aviation Safety Authority (CASA)~\cite{casa_drones} in Australia have established frameworks incorporating risk-based drone classification, cybersecurity protocols, and operational licensing to ensure the safe integration of UAVs into the national airspace. In India, the Digital Sky Platform, governed by the Directorate General of Civil Aviation (DGCA)~\cite{dgcaDigitalSky}, ensures compliance with the established rules and prevents unauthorized drone operations through automated flight authorization, real-time tracking, and geo-fencing~\cite{srivastava2020review}. 

Despite the effective steps taken by the international aviation authorities, there is a lack of universal global regulations that apply to all countries, and there exists inconsistencies in enforcement rules. Another substantial challenge in drone operations is a lack of mechanisms for dynamic policy updates, e.g., concerning {\em No-Fly Zones (NFZs)}. Dynamic NFZs may be required for the security of the local events. The current regulations only focus on basic guidelines such as height and payload restrictions in static NFZs, and the existing drone systems are not designed such that an authorized agency can enforce the dynamic NFZs.  
\par
To ensure UAV security and regulatory compliance, we envision that a centralized UAV regulatory authority must be established, as shown in Figure \ref{fig:considered_system}. This regulator can oversee drone registration, secure flight permissions, real-time tracking, and cybersecurity enforcement. Secure communication between UAVs and the regulator can be achieved through conventional protocols based on Public Key Infrastructure (PKI) and zero-trust security models, preventing unauthorized manipulation of drone operations. The regulator can also implement remote identification mechanisms, intrusion detection systems, and blockchain-based logging to track drones and enforce compliance. Additionally, AI-based risk assessment, cybersecurity certification for UAV manufacturers, and compliance audits can strengthen UAV security frameworks. Governments can ensure safe UAV integration, prevent unauthorized drone access, and enhance national security by establishing a dedicated regulatory authority with stringent cybersecurity mandates.
\par


\subsection{Potential Research Problems}
The above discussion highlights the potential role of the regulator in ensuring secure and compliant drone operations. To design and develop a novel framework for equipping drones with a regulatory-compliant system, as shown in Figure \ref{fig:considered_system}, we must address the following key research questions:
\begin{enumerate}[noitemsep]
    \item \textbf{How would a drone system be designed such that it must follow every command from a regulatory authority?}
    The regulatory authority should be able to send a command (e.g., ``return to home'' command) to a maliciously behaving drone, and the drone should have no option but to follow the command. 
    \item \textbf{How will the regulatory authority securely communicate with the drone for sending the command?}
    The communication between the drone and the regulatory authority should be established and secured to restrict unauthorized users from sending malicious commands to the drone.
    
    \item \textbf{How would the regulatory authority detect whether the drone is behaving maliciously or not?}
    The regulatory authority should have a mechanism to establish the correct functionality of the drone.
\end{enumerate}

\subsection{Potential Solutions}

Some potential measures for establishing a robust Regulatory Authority (RA) in UAV operations include:
\\
\noindent
\textbf{Trusted Execution Environment (TEE):} A drone can potentially be equipped with a TEE which may ensure that commands sent by the regulatory authority and approved by the drone-TEE must be processed by the software stack of the drone.
\\
\noindent
\textbf{Real-time Monitoring of No-Fly Zones (NFZs):} The regulatory authority must be enabled to perform continuous monitoring and dynamic enforcement of NFZs to prevent unauthorized UAV access to restricted areas.
\\
\noindent
\textbf{Standardized cybersecurity policies:} Developing and enforcing global cybersecurity policies to regulate UAV operations across different jurisdictions.
\\
\noindent
\textbf{Intrusion Detection and Prevention Systems (IDPS):} Implementing AI-driven IDPS for UAV networks to detect anomalous behavior and prevent cyber threats in real-time. 
\par
The regulatory authority should be able to detect the maliciously behaving drone, and after post detection of the malicious behavior of the drone, the drone should be able to receive a command from the law enforcement or the regulatory authority and should be able to follow it strictly without violating it. 
This research introduces a novel direction in drone security by integrating policy enforcement into trusted hardware, offering a robust, scalable, and policy-aligned framework for secure and compliant drone operations.
\section{Conclusion}\label{sec8_Conclusion}
In this paper, we have presented novel taxonomies for analyzing the threats and countermeasures in the field of UAVs. We have summarized and tabulated various possible attacks using the threat taxonomy, and evaluated the impact of available defenses using the countermeasure taxonomy. Our analysis reveals the existing limitations and challenges in securing drone networks, motivating us to explore the potential of a regulated drone network. Hence, we have examined the need for regulations in the drone networks and explored the problems that could be resolved for the establishment of the regulatory authority in the UAV operations to maintain security and privacy.

\bibliographystyle{unsrtnat}
\bibliography{reference}

\end{document}